\documentclass{article}

\usepackage{arxiv}

\usepackage[utf8]{inputenc} % allow utf-8 input
\usepackage[T1]{fontenc}    % use 8-bit T1 fonts
\usepackage{hyperref}       % hyperlinks
\usepackage{url}            % simple URL typesetting
\usepackage{booktabs}       % professional-quality tables
\usepackage{amsfonts}       % blackboard math symbols
\usepackage{graphicx}
\usepackage{grffile,float,xcolor,xspace}
\usepackage{amssymb,amsfonts,amsmath,graphicx}
\usepackage[english]{babel}
\usepackage{subfigure}

\title{Influence of Allee Effect on Extreme Events 
in Coupled Three Species Systems}

\author{Deeptajyoti Sen \\
	Department of Physical Sciences\\
	IISER Mohali\\
	Manauli, Punjab - 140306 \\
	\texttt{deeps.sen.25@gmail.com} \\
	\And
	Sudeshna Sinha\thanks{Corresponding author} \\
	Department of Physical Sciences\\
	IISER Mohali\\
	Manauli, Punjab - 140306 \\
	\texttt{sudeshna@iisermohali.ac.in} \\
}

\date{}

\renewcommand{\shorttitle}{Influence of Allee Effect on Extreme Events}

\begin{document}
\maketitle

\begin{abstract}
	We consider the dynamics of two coupled three-species population patches, incorporating the Allee Effect, focussing on the onset of extreme events in the coupled system. 
%At the outset we review our earlier results on the enhancement of extreme events through the Allee effect. 
First we show that the interplay between coupling and the Allee effect may change the nature of the dynamics, with regular periodic dynamics becoming chaotic in a range of Allee parameters and coupling strengths. Further, the growth in the vegetation population displays an explosive blow-up beyond a critical value of coupling strength and Allee parameter. Most interestingly, we observe that beyond a threshold of coupling strength and Allee parameter, the population densities of all three species exhibit non-zero  probability of yielding extreme events. The emergence of extreme events in the predator populations in the patches is the most prevalent, and the probability of obtaining large deviations in the predator populations is not affected significantly by either the coupling strength or the Allee effect. In the absence of the Allee effect the prey population in the coupled system exhibits no extreme events for low coupling strengths, but yields a sharp increase in extreme events after a critical strength of coupling. The vegetation population in the patches display a small finite probability of extreme events for strong enough coupling, only in the presence of Allee effect. Lastly we consider the influence of additive noise on the continued prevalence of extreme events. Very significantly, we find that noise suppresses the unbounded vegetation growth that was induced by a combination of Allee effect and coupling. Further, we demonstrate that noise mitigates extreme events in all three populations, and beyond a noise level we do not observe any extreme events in the system at all. This finding has important bearing on the potential observability of extreme events in natural and laboratory systems.
\end{abstract}

% keywords can be removed

\section{Introduction}

Investigating the advent of extreme events, signalling behaviour beyond normal variability in the dynamics of complex systems, has enormous relevance from the viewpoint of basic understanding of complex systems, as well as implications for risk assessments from catastrophic surges \cite{powergrid1,weather,optical,powergrid2}. So exploring the emergence of such events in models and real-world systems, as well as the search for mechanisms and processes that may underlie extreme events, has witnessed much research interest in recent years \cite{extreme}. An extreme event can be labelled as one where a state variable displays very large, relatively rare, fluctuations from the average value. That is, in the course of its evolution the system exhibits occasional uncorrelated excursions that are significantly different from the mean. 
%Additionally, these large deviations are recurrent, but rare, and occur aperiodically. 
So the most commonly employed signature of extreme events in phenomena ranging from oceanography \cite{ocean} to financial markets \cite{market}, is uncorrelated recurrent deviations larger than a prescribed threshold of typically 3-8 standard deviations away from the average value.

A very important direction in the study of extreme events is to unearth generic mechanisms that can give rise such large deviations in the dynamics. While extreme events in stochastic models has been extensively studied over decades \cite{santhanam,satya1}, the advent of extreme events in deterministic dynamical systems, without intrinsic or extrinsic stochasticity, has garnered focus in recent times \cite{ulrike1,balki1,ulrike2,ulrike3,promit}.
Focussing on this direction, we consider emergent extreme phenomena in patches of  vegetation-prey-predator systems coupled through Lotka-Volterra interactions, incorporating the biologically important Allee effect. In population dynamics the Allee effect reflects the advantageous influence of conspecific interactions on population growth \cite{Courchamp2008,Dennis1989,Sen2021}, and captures the impact of small population size on the long-term persistence of a population. Further we consider  
%we investigate the effect of additive noise in the system, focusing on 
the role of noise on the propensity of extreme events in this coupled three-species system. Beyond modelling population dynamics in ecosystems, these results have broad bearing on the mechanisms that can enhance extreme events in deterministic dynamical systems, and the effect of stochasticity on their prevalence.

%This paper spans a brief review of earlier results and then goes on to new results that extend the past results. 
Our paper is organized as follows. At the outset, in Section 2 we introduce the model of three interacting species, incorporating the Allee effect, and recall some significant results in this system.
%: first, Allee effect induces chaos in this system. Further, when Allee effect is sufficiently strong, vegetation grows explosively. Most importantly, Allee effect enhances extreme events in this three-species chain. We also recall the interesting observation that noise suppresses the unbounded blow-up of vegetation, and also subdues the extreme events induced by Allee effect. This suggests a natural mechanism that can potentially mitigate extreme events in population chains. 
In Section 3 we go on to explore the dynamics of coupled patches of such three-species systems to establish the generality and broad scope of our findings. Lastly, in Section 4, we summarize our results and discuss their potential implications.

%\section{Three-species food chain model incorporating the  Allee Effect}
 
\section{Three species food chain model with the Allee effect}

The local and global dynamics of three species interacting models has significant impact in complex system research, in particular in theoretical ecology. Here we will consider a vertical food chain model incorporating the dynamics of the snowshoe hare and the Canadian lynx populations, based on observed data. This model consists vegetation (denoted by $u$), prey (denoted by $v$) and predator (denoted by $w$) and also incorporates  the Allee effect into the growth of predator. The dynamics of the model can be described by the following coupled nonlinear ordinary differential equations:

\begin{equation}
 \label{eq:finalmodel}
     \begin{split}
         \dot{u} &= f(u,v,w)\,=\,a u \ - \ \alpha_{1}f_{1}(u,v),\\
         \dot{v} &= g(u,v,w)\,=\,\alpha_{1}f_{1}(u,v) \ A(v) \ - \ bv \ - \ \alpha_{2}f_{2}(v,w),\\
         \dot{w} &= h(u,v,w)\,=\,\alpha_{2}f_{2}(v,w) \ - \ c(w-w^{*}),
     \end{split}
 \end{equation}

\noindent where $a$, $b$ and $c$ are the growth rates of vegetation, prey and predator respectively. Here we consider the interaction between vegetation and prey follow Holling type II functional response $f_{1}(u,v)\,=\,\frac{uv}{1+ku}$ whereas the interaction between prey and predator is considered to follow Lotka-Volterra type interaction, described by $f_{2}(u,v)\,=\,u v$. The parameter $k$ corresponds to the average time spent for processing a food, and is termed ``handling time''. Here $\alpha_{1}$ denotes the maximum growth rate of prey, which is the product of the ingestion rate and a constant factor less than unity, considering the fact that not all ingested vegetation population converted into prey biomass. The parameter $\alpha_{2}$ is the parameter that corresponds to $\alpha_{1}$ in the predator population. The predator is considered to be generalist, accounting for the fact that it can persist at an equilibrium $w^{*}$, either in absence of prey or when it's concentration is low. Additionally, we incorporate the Allee effect into the prey's growth by introducing the following functional form

$$A(v)\,=\,\frac{v}{v+\theta},$$

\noindent where $\theta$ is the Allee parameter reflecting the critical prey density at which the probability of successful mating would be half. This form is characteristic of the fact that the per-capita reproduction rate becomes smaller at low prey density. This kind of Allee effect appears due to lack of mating partners, low fertilization efficiency, cooperative breeding mechanism, etc in the context of biology. The dynamics of this three species model and the consequences of the Allee effect on this system has been very recently studied in \cite{scirep_recent}. We recall the principal results from that study below:

\begin{itemize}
    \item The Allee effect induces explosive increase (which we term a ``blow-up'') of the vegetation population, i.e. there is a critical threshold of the Allee parameter beyond which the vegetation population has a positive probability of unbounded growth.
    \item While this three species system is regular when the Allee effect is absent or small, sufficiently large Allee effect induces chaos in the system.
    \item The Allee effect also has an impact on the development of extreme events in the three-species system, with the Allee effect typically enhancing the probability of obtaining such events.
    \item Lastly, additive noise in this three-species system mitigates the blow-up of the vegetation population, as well as suppresses extreme events. 
\end{itemize}

\section{Patches of three-species systems coupled through cross-predation}

In order to gauge the generality of the phenomena observed in a single patch in our earlier work, we now explore the dynamics of two coupled patches, where each patch again has three species, namely vegetation, prey and predator. The populations in local patches are connected in such a way that  predator of one patches can attack the prey of neighbouring patches and vice versa. 
%That is, the patches are coupled via cross predation, with the form of the prey-predator interaction between the neighbouring patches following the Lotka-Volterra functional form. Importantly again, our model incorporates the Allee effect in each patch. 
%Specifically, we consider the earlier three species population model consisting vegetation, prey and predator populations, distributed over two patches. 
%These two patches are coupled through in such a way that predator of one patch can attack the prey of another patch and vise versa. 
This coupling strategy, known as coupling through cross-predation, signifies that predators are more mobile compared to prey and can move into another patch to capture the prey. Dynamics of the coupled three species system can be described by the following sets of equations:

\begin{equation}
\label{eq:coupled_system}
    \begin{split}
        \dot{u}_{1} &= f(u_{1},v_{1},w_{1}),\\
        \dot{v}_{1} &= g(u_{1},v_{1},w_{1}) \ - \ C \ v_{1}w_{2},\\
        \dot{w}_{1} &= h(u_{1},v_{1},w_{1}) \ + \ C \ v_{2}w_{1},\\
         \ & \ \\
        \dot{u}_{2} &= f(u_{2},v_{2},w_{2}),\\
        \dot{v}_{2} &= g(u_{2},v_{2},w_{2}) \ - \ C \ v_{2}w_{1},\\
        \dot{w}_{2} &= h(u_{2},v_{2},w_{2}) \ + \ C \ v_{1}w_{2}.
    \end{split}
\end{equation}

\medskip

\noindent Here the populations of vegetation, prey and predator in $i$-th patch ($i\,=\,1,2$) are denoted by $u_{i}$, $v_{i}$ and $w_{i}$ respectively and $C$ is a parameter reflecting the inter-patch coupling strength. The function $f_{i}, g_{i}, h_{i},\,i\,=\,1,2$ have the same form as in a single patch given by Eqn.~\eqref{eq:finalmodel}, with parameters $a_{i}, b_{i}, c_{i}, \alpha_{1i}, \alpha_{2i}, \theta_{i},\,i\,=\,1,2$. To start our analysis we assume that $a_{1}\,=\,a_{2}\,=\,a,\, b_{1}\,=\,b_{2}\,=\,b,\,c_{1}\,=\,c_{2}\,=\,c,\,\alpha_{11}\,=\,\alpha_{12}\,=\,\alpha_{1},\,\alpha_{21}\,=\,\alpha_{22}\,=\,\alpha_{2},\,\theta_{1}\,=\,\theta_{2}\,=\,\theta$. Although considering identical parameters is not accurate from the ecological point of view, it provides a good test bed for investigations and serves as an useful starting point for analyzing the coupled system. In this study we consider the parameter values  $a\,=\,1$, $b\,=\,1$, $c\,=\,10$, $w^{*}\,=\,0.006$, $\alpha_{1}\,=\,0.5$, $\alpha_{2}\,=\,1$, $k\,=\,0.05$ \cite{blasius}. We explore the dynamics of the coupled system under varying Allee parameter $\theta$ and coupling strength $C$, through numerical simulations using the Runge-Kutta fourth order algorithm. We have corroborated the stability and convergence of our results with respect to decreasing step size.

\begin{figure}
    \centering
    \includegraphics[width=0.5\textwidth]{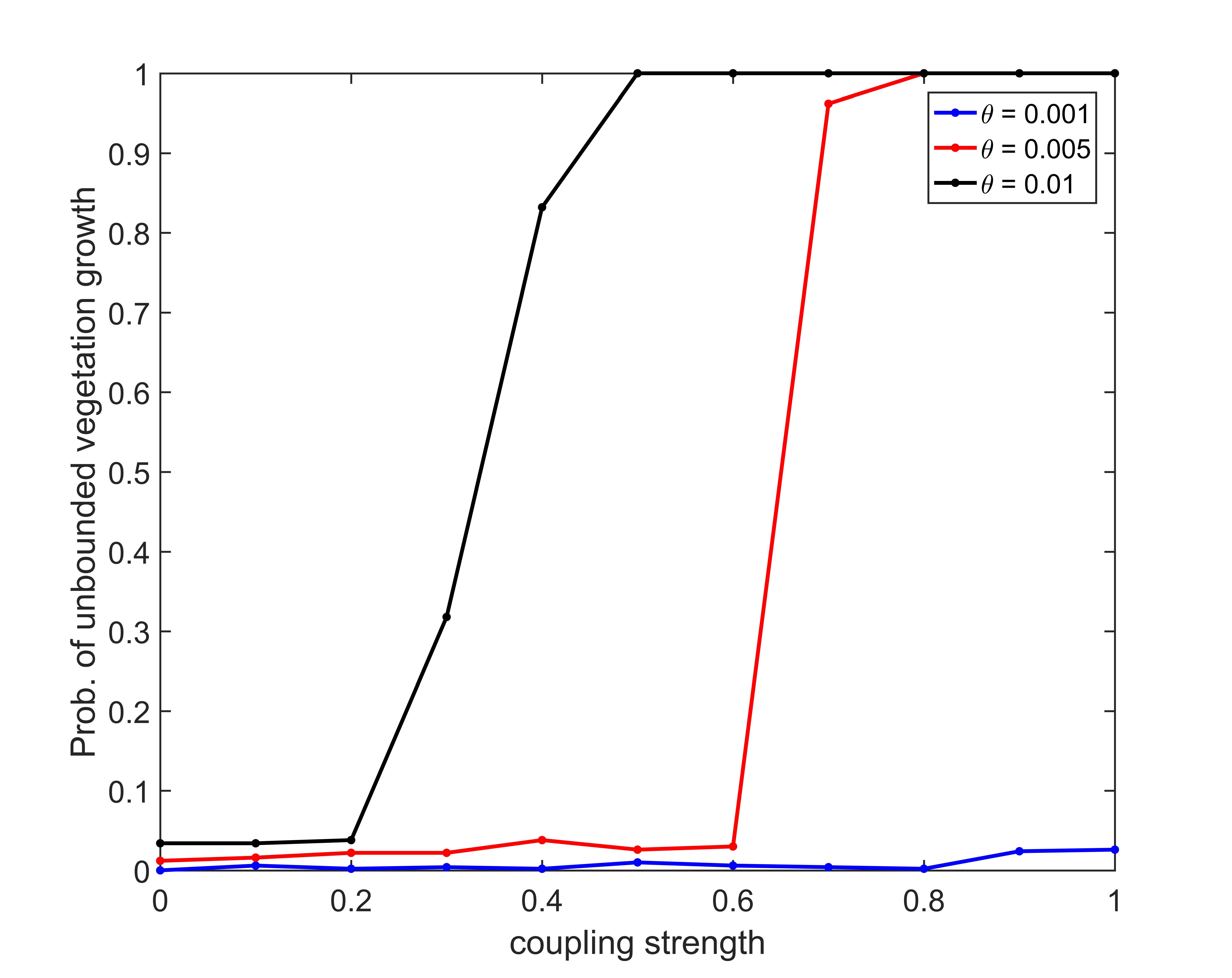}
    \caption{Probability of unbounded vegetation growth in coupled patches with respect to the coupling strength $C$. A blow-up to be considered to occur when the vegetation population in any patches exceeds $10^3$. Three different values of $\theta$ are considered and the probability is estimated from a sample of 500 initial states distributed randomly in a hyper-cube ($u_{i}\in[0:4]$, $v_{i}\in[0:2]$, $w_{i}\in[0:5]$) over each patch.} 
    \label{fig:Prob_blowup_coupledpatch}
\end{figure}

\section*{Temporal evolution of population densities 
%and phase attractors 
in the coupled patches}

Our first significant observation is the emergence of ``blow-ups'' in the vegetation population densities beyond a threshold of coupling strength. This threshold decreases with increasing the Allee parameter $\theta$, namely the onset of unbounded vegetation growth occurs at weaker coupling for stronger Allee effect. In order to quantify the advent of such blow-ups, we estimate the probability of unbounded vegetation growth from many initial states followed over an extended period of time, ensuring that the estimated values are converged with respect to sample size of initial conditions. In Fig.~\ref{fig:Prob_blowup_coupledpatch} we display the results thus obtained, for varing coupling strengths $C$, for values of the Allee parameter $\theta\,=\,0.001,\,0.005,\,0.01$. It is clearly noticeable from the figure that for each value of the Allee parameter $\theta$, there exists a critical value $C^*_{\theta}$ of coupling strength, beyond which vegetation has a nonzero probability of blow-up. It is also evident that for increasing values of $\theta$, the value of $C^*_{\theta}$ reduces. This clearly demonstrates that the Allee effect enhances the propensity of explosive vegetation growth, a trend that is consistent with results from a single patch.

\begin{figure}[!ht]
\centering
%\mbox{\subfigure[]{\includegraphics[width=0.48\textwidth]{TimeSeires_Patch1_theta0dot003_c0}}
%\quad
%\subfigure[]{\includegraphics[width=0.48\textwidth]{PhasP_patch1_theta0dot003_c0}}}
%\mbox{\subfigure[]{\includegraphics[width=0.48\textwidth]{TimeSeires_Patch1_theta0dot003_c0dot4}}
%\quad
%\subfigure[]{\includegraphics[width=0.48\textwidth]{PhasP_patch1_theta0dot003_c0dot4}}}
%\mbox{\subfigure[]{\includegraphics[width=0.48\textwidth]{TimeSeires_Patch1_theta0_c0dot5}}
%\quad
%\subfigure[]{\includegraphics[width=0.48\textwidth]{PhasP_patch1_theta0_c0dot5}}}
%\mbox{\subfigure[]{\includegraphics[width=0.48\textwidth]{TimeSeires_Patch1_theta0dot003_c0dot5}}
%\quad
%\subfigure[]{\includegraphics[width=0.48\textwidth]{PhasP_patch1_theta0dot003_c0dot5}}}
\mbox{\subfigure[]{\includegraphics[width=0.48\textwidth]{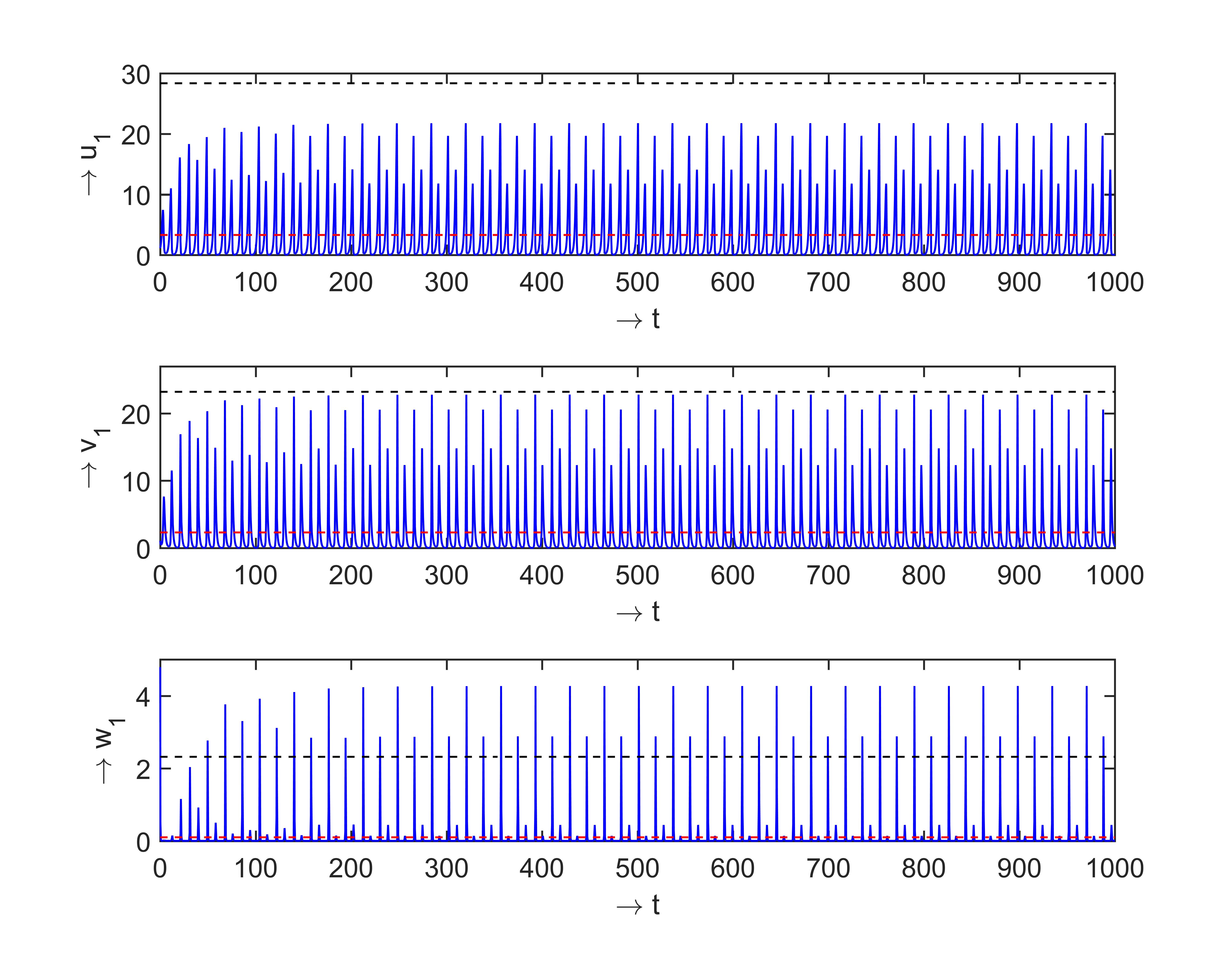}}
\quad
\subfigure[]{\includegraphics[width=0.48\textwidth]{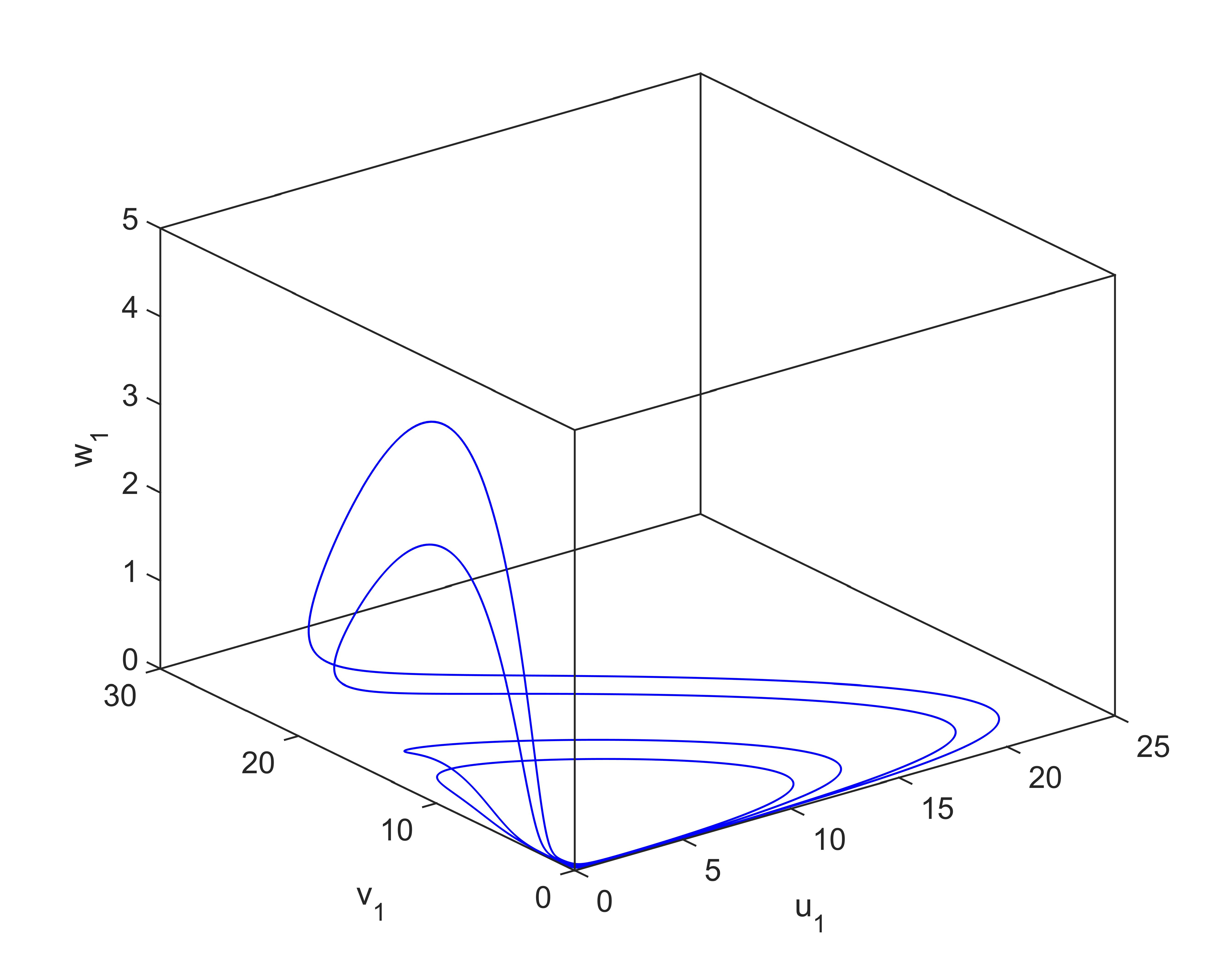}}}
\mbox{\subfigure[]{\includegraphics[width=0.48\textwidth]{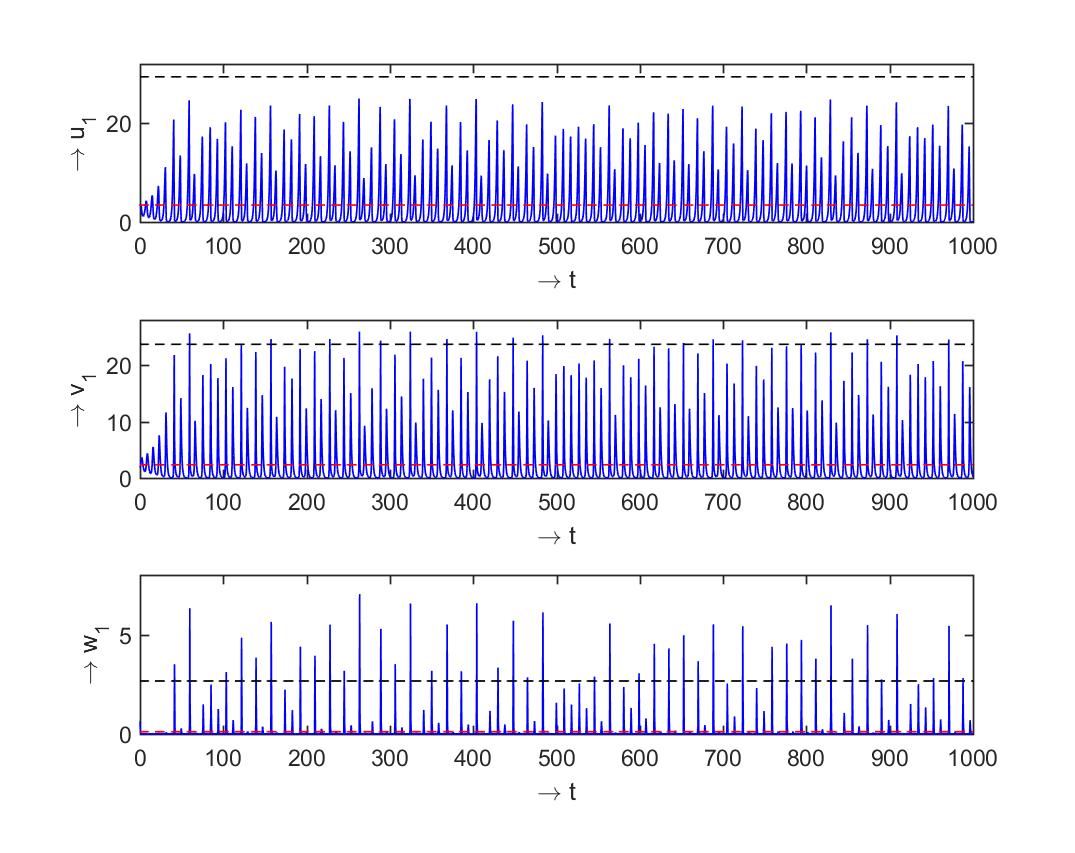}}
\quad
\subfigure[]{\includegraphics[width=0.48\textwidth]{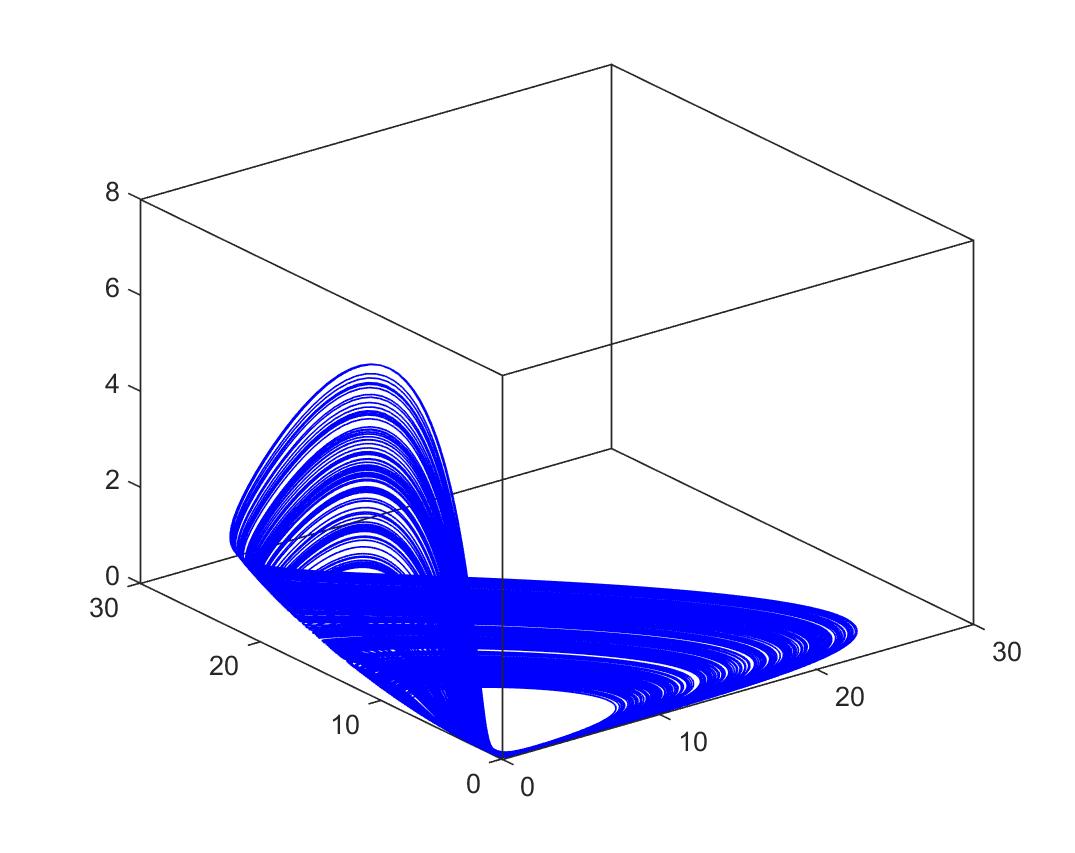}}}
\caption{Time series for the vegetation, prey and predator populations in a patch, and the corresponding phase space attractor, of the system given by Eqns.~\eqref{eq:coupled_system}, for coupling strength $C=0.1$, and the Allee parameter $\theta$ equal to $0$ (a-b) and $0.003$ (c-d). 
%Left panel shows the time series for the vegetation, prey and predator populations in a patch, for coupling strength (top) 
%$c\,=\,0$, (b) $c\,=\,0.4$, and (bottom) $c\,=\,0.8$, and  the right panel displays the corresponding the attractor. 
The red dashed line corresponds to mean $\mu$ of the time series and the black dashed line indicates the $\mu + 5\sigma$ threshold.}
\label{fig:timeseries&phasep_with_coupling}
\end{figure}

\begin{figure}[!ht]
    \centering
    \includegraphics[width=0.3\textwidth]{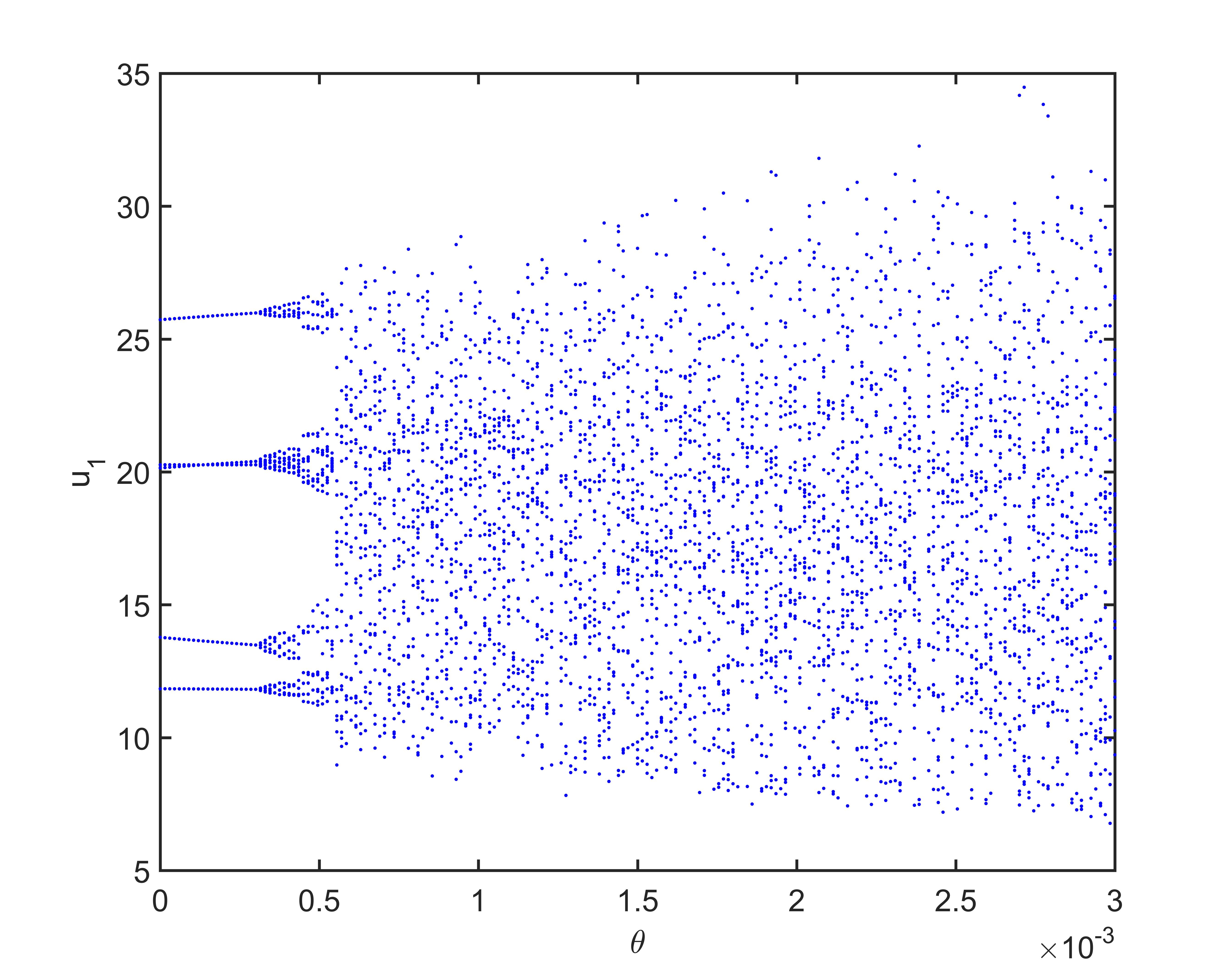}
\includegraphics[width=0.3\textwidth]{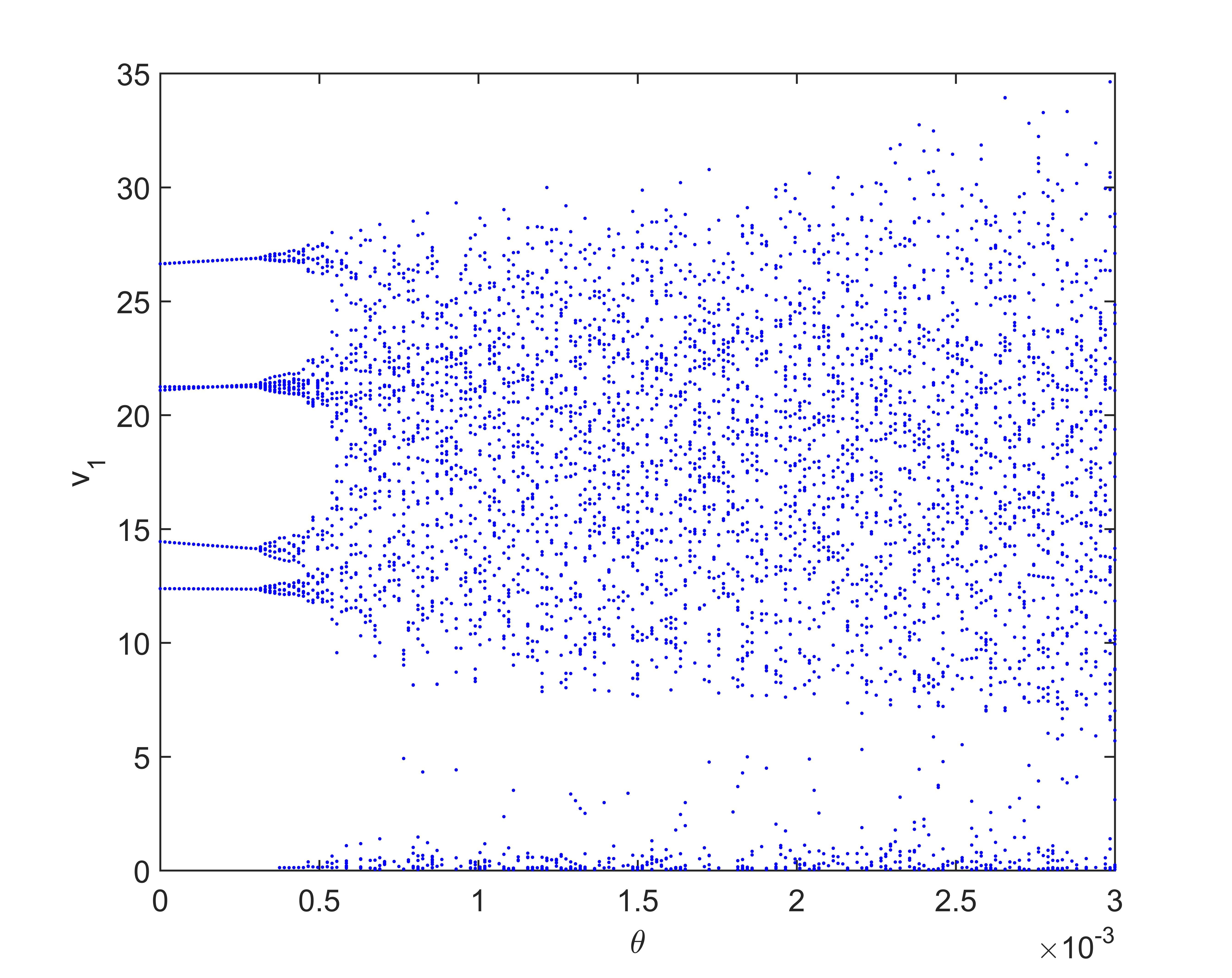}
\includegraphics[width=0.3\textwidth]{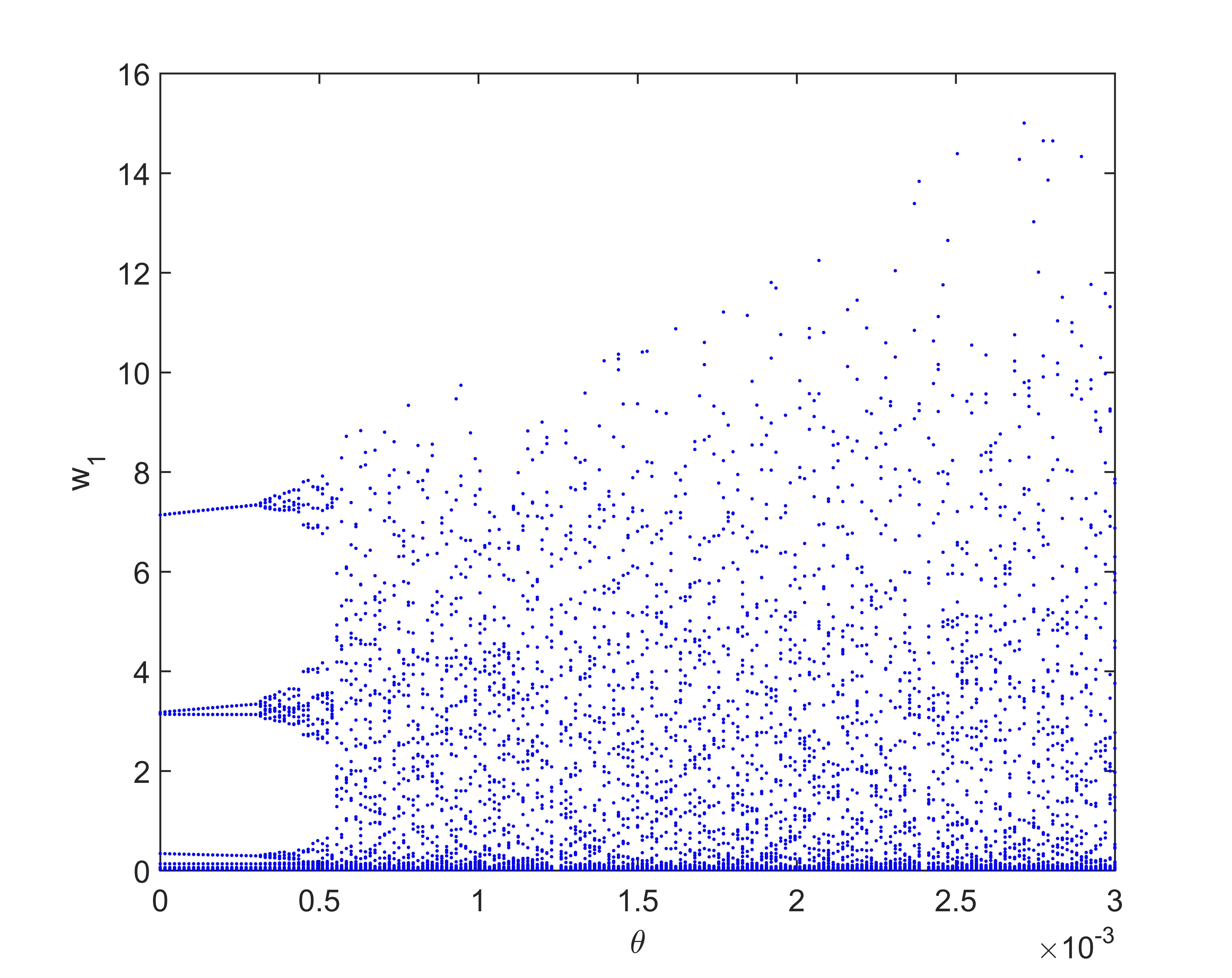}
    \caption{Bifurcation diagram of (left to right) vegetation, prey and predator population densities, with respect to the Allee parameter $\theta$ for the coupling strength $C\,=\,0.5$.}
    \label{fig:Bfdiag_theta_c0dot5}
\end{figure}

Next we look into the temporal evolution of the population densities and the emergent attractors in phase phase, for the coupled patches. To illustrate the dynamics of the coupled system we present two representative time series and the corresponding phase space attractors in Fig.~\ref{fig:timeseries&phasep_with_coupling}, for Allee parameter $\theta\,=\,0$ and $\theta\,=\,0.003$, with coupling strength $C=0.1$. 
%and Fig.~\ref{fig:timeseries&phasep_with_coupling_theta0dot003} respectively by varying the coupling strength $C$. 
It is clearly evident that Allee effect induces chaos in the coupled system, and we obtain chaotic attractors for sufficiently large $\theta$. This phenomenon is demonstrated more rigorously through the bifurcation diagram in Fig.~\ref{fig:Bfdiag_theta_c0dot5} which displays the emergence of chaos after a critical value of $\theta$. This trend is similar to that observed in a single patch. So the influence of the Allee effect observed in a single patch generalizes to two coupled patches. Further, importantly, increasing the Allee effect parameter $\theta$ increases the size of the chaotic attractor in the coupled system. The sparse points at the high values of $u$, $v$ and $w$ are manifested as extreme events in the time series.  

\begin{figure}[!ht]
    \centering
\includegraphics[width=0.3\textwidth]{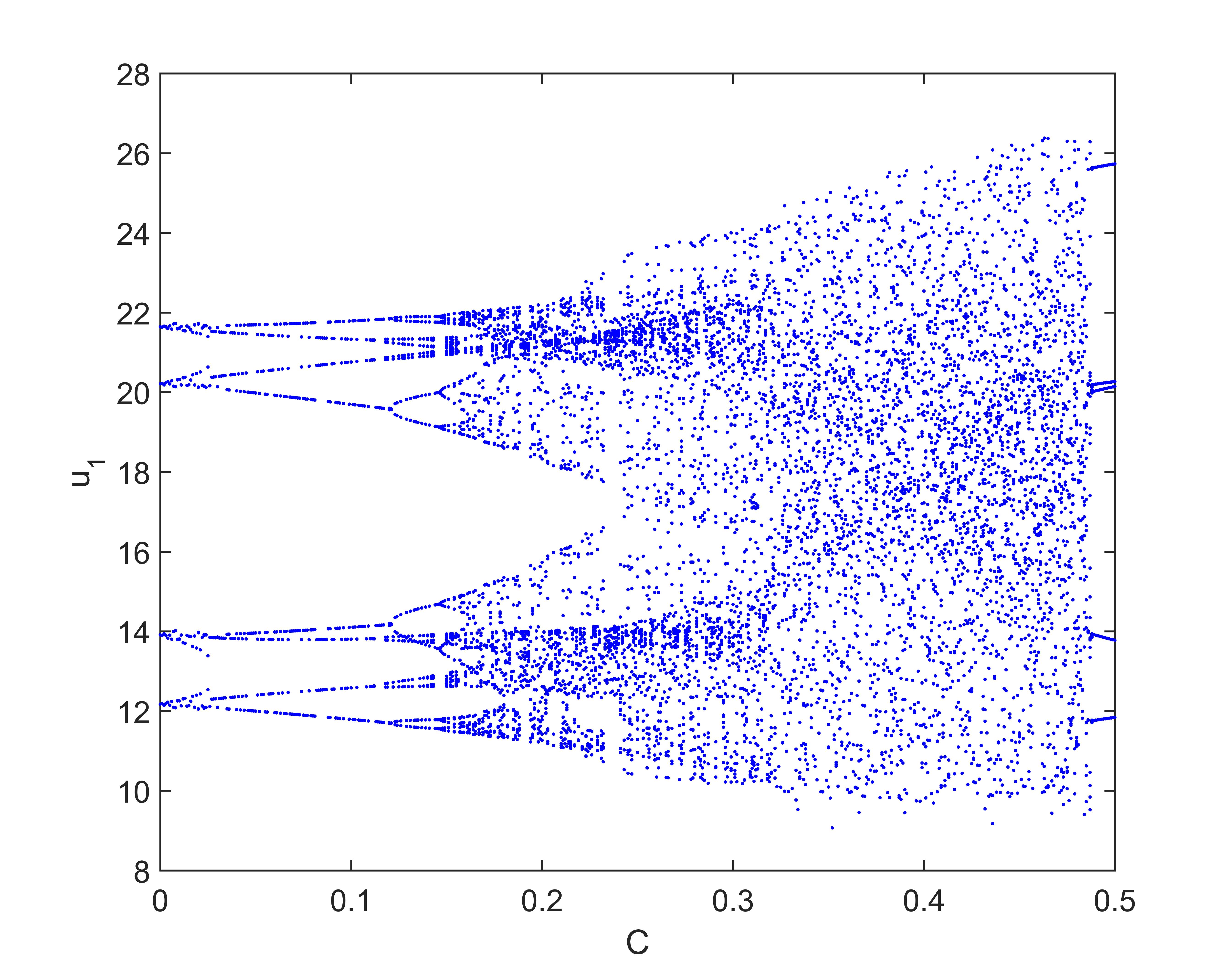}
\includegraphics[width=0.3\textwidth]{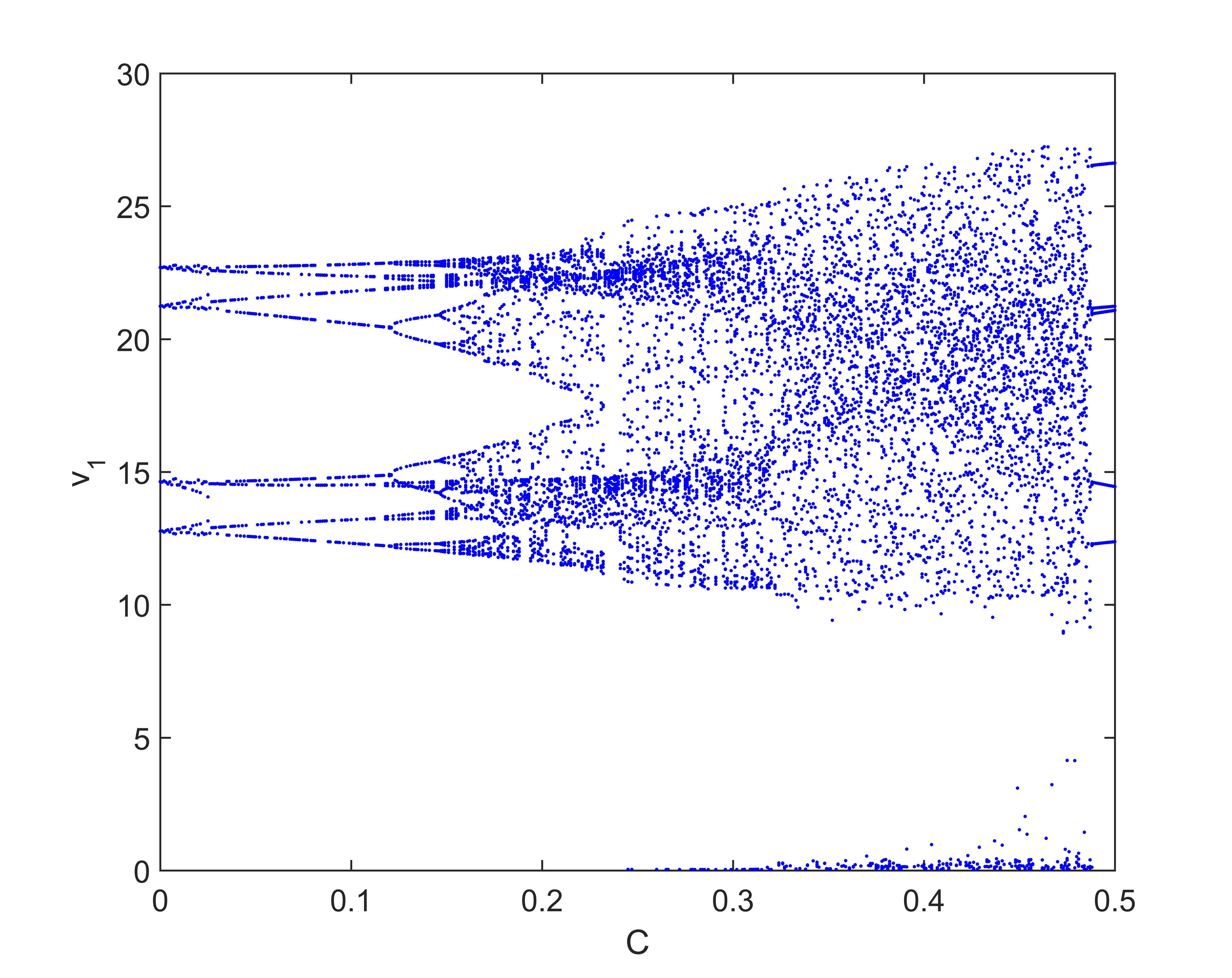}
\includegraphics[width=0.3\textwidth, height=1.75in]{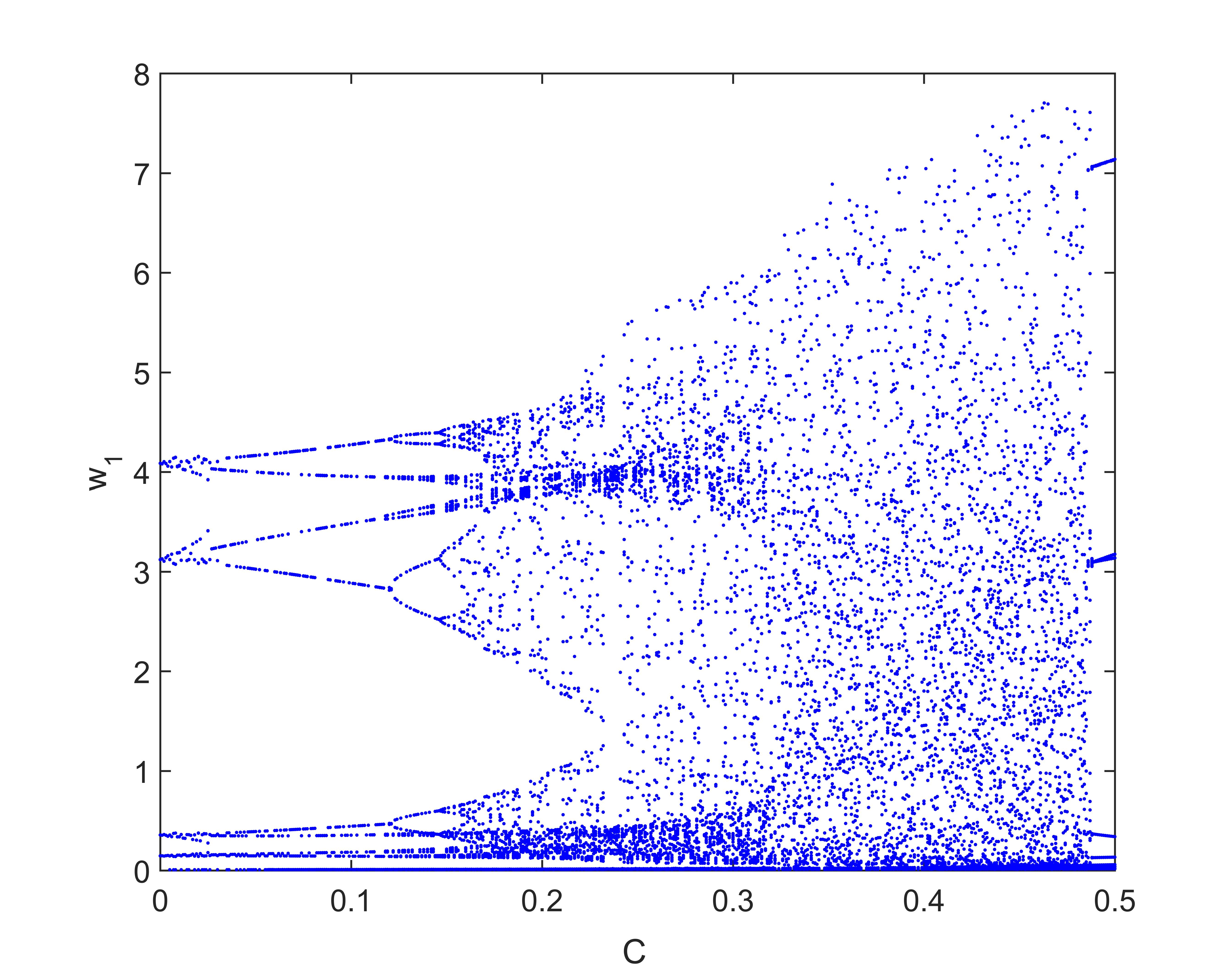}
\includegraphics[width=0.3\textwidth]{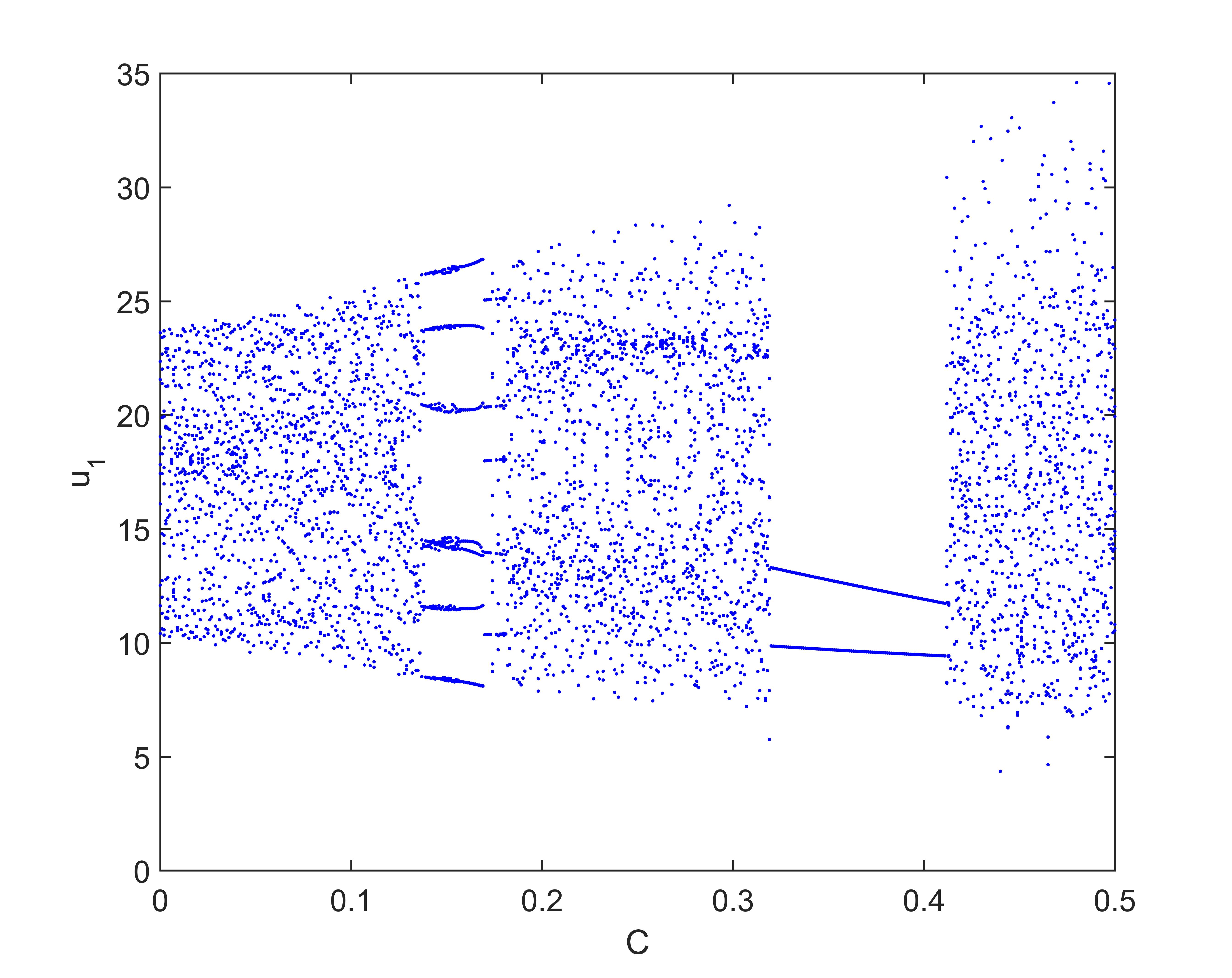}
\includegraphics[width=0.3\textwidth]{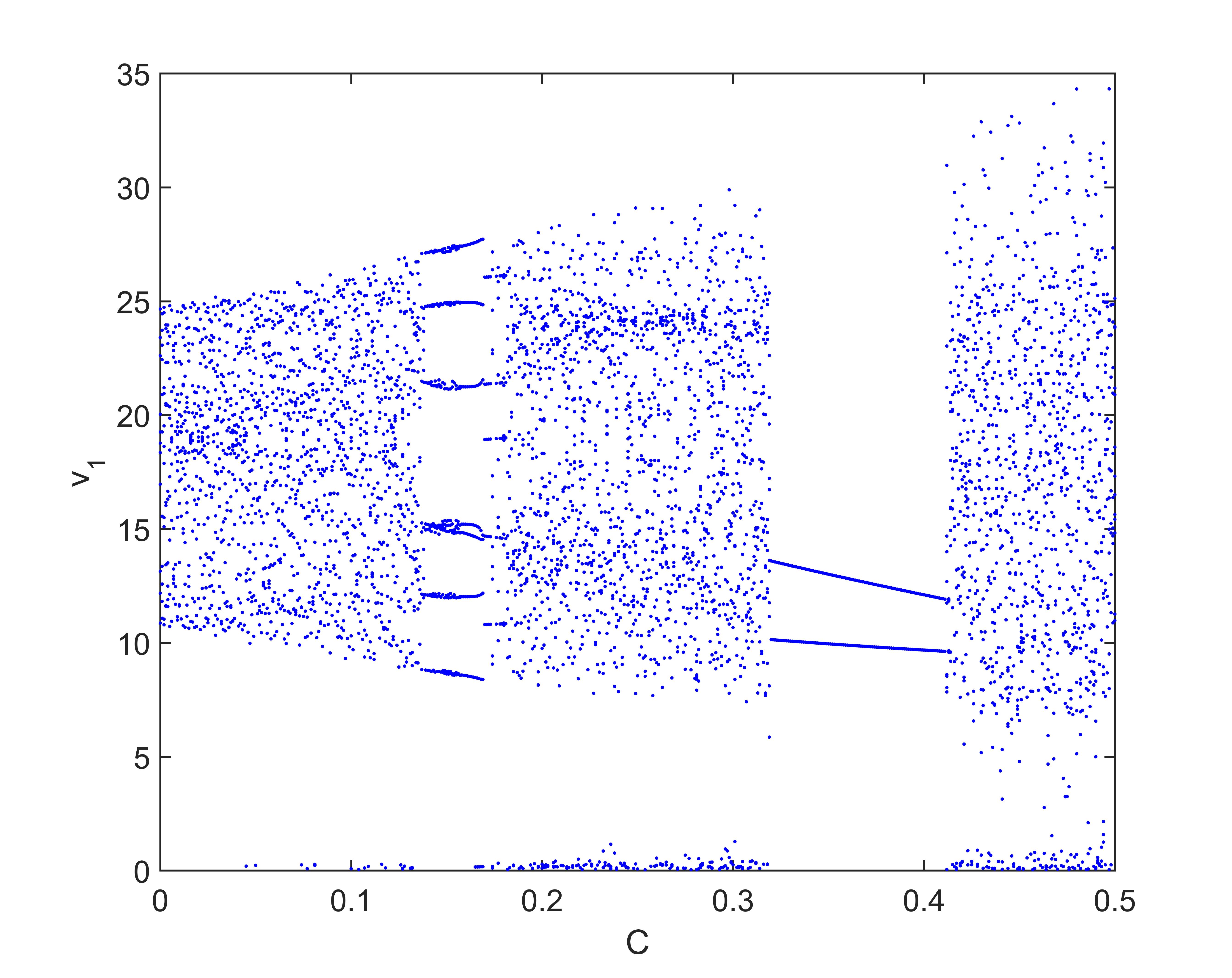}
\includegraphics[width=0.3\textwidth, height=1.5in]{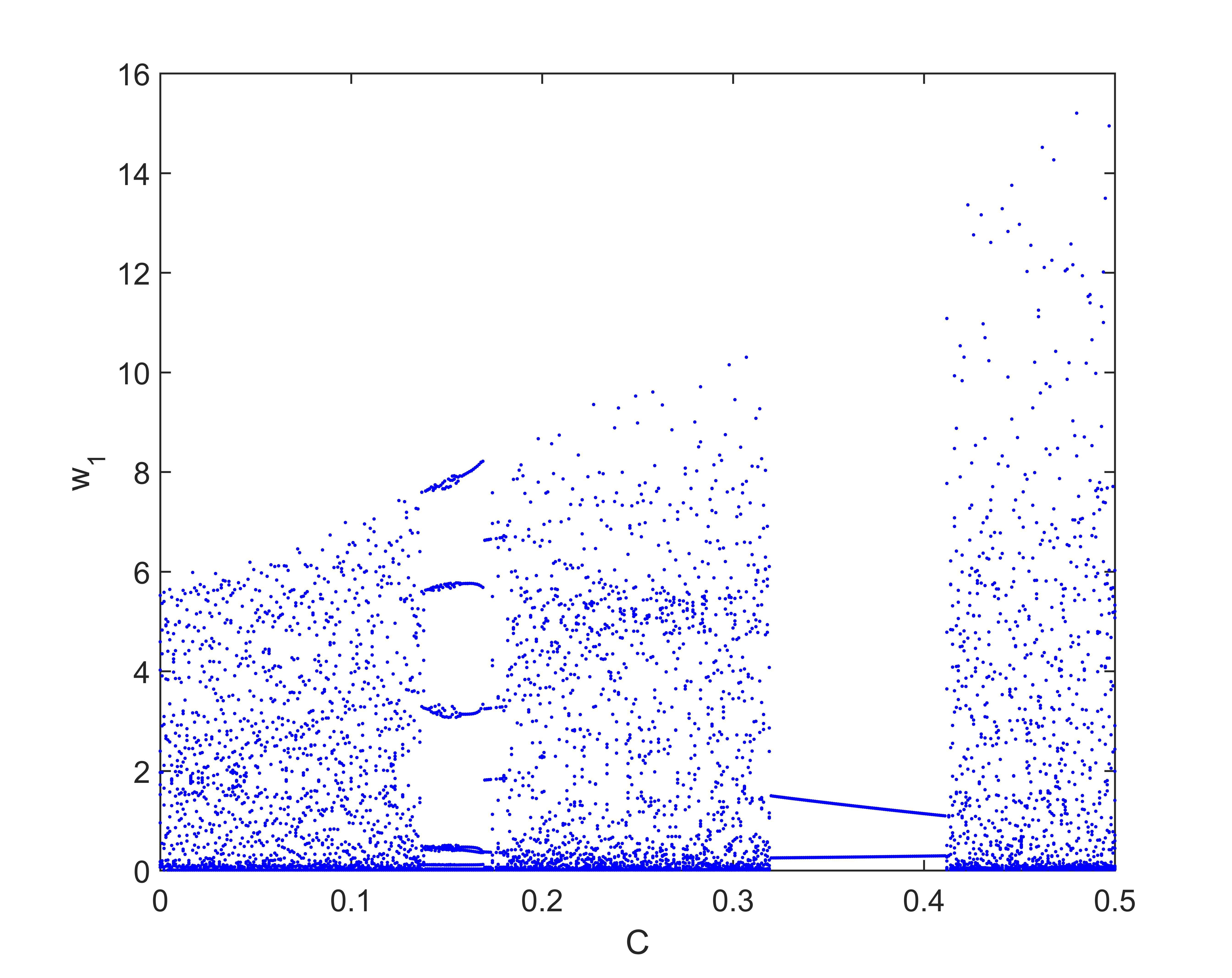}
    \caption{Bifurcation diagram of (left to right) vegetation, prey and predator populations in a patch, with respect to the coupling strength $C$, for (top row) $\theta\,=\,0$, and (bottom row) $\theta\,=\,0.003$.}
    \label{fig:Bfdiag_wrt_coupling}
\end{figure}

Fig.~\ref{fig:Bfdiag_wrt_coupling} shows the bifurcation sequence with respect to coupling strength $C$, for $\theta = 0$ and $\theta= 0.003$. It has already been observed that in a single patch when the Allee effect is absent, all population densities evolve periodically with the system attracted to a period-$4$ orbit.
%which is consistent with the result $c\,=\,0$, $\theta\,=\,0$ in Fig.~\ref{fig:timeseries&phasep_with_coupling_theta0}a \& Fig.~\ref{fig:timeseries&phasep_with_coupling_theta0}b. But for increasing the coupling strength $C$ ($C\,=\,0.4$), 
However, interestingly, under coupling, even in the absence of Allee effect, we observe that the populations in the patches evolve aperiodically when coupling is sufficiently strong (see Fig.~\ref{fig:Bfdiag_wrt_coupling}, top row). One also observes a periodic window arising near $C\,=\,0.5$ in the bifurcation diagram as a result of interior crisis. 
%Similar trends of chaotic regimes, interspersed with periodic windows, can be seen in the bifurcation diagrams when the Allee parameter is non-zero (Fig.~\ref{fig:Bfdiag_wrt_coupling}, bottom row). The notable difference arising from the Allee effect, is that there is chaos for low coupling strengths, including the case of $C=0$ (i.e. the uncoupled case) for finite $\theta$, while weakly coupled patches with no Allee effect exhibits regular dynamics. 
 
The bifurcation diagrams for the case of Allee parameter $\theta=0.003$ is displayed in Fig.~\ref{fig:Bfdiag_wrt_coupling} (bottom row). The first significant observation is that chaos arises in the presence of Allee effect, even in uncoupled patches. Further we observe that increasing the coupling strength typically increases the size of the chaotic attractor, except in a small range ($C \in [0.32,0.411]$) where the system settles into a small two-periodic attractor as a result of interior crisis. This two-periodic attractor becomes unstable once again as the coupling strength increases, giving rise to the sudden emergence of a large chaotic attractor. So comparing the bifurcation scenarios for the two different cases we can conclude that weak coupling induces chaos in the presence of Allee effect, while for strong coupling chaos arises even in the absence of Allee effect.

Further, we explore the synchronization between the populations of vegetation, prey and predator in the two patches. In order to quantify the degree of synchronization we calculate the average synchronization error, defined as the mean square difference of the corresponding population densities of the two patches, obtained by averaging over long time and many initial states. It is clear that there is {\em no synchronization}, even for strong coupling. The closest synchronization (i.e. lowest synchronization error) is achieved in the parameter window supporting a period-2 orbit for $\theta=0.003$, but typically the patches do not synchronize irrespective of the presence or absence of the Allee effect.

\begin{figure}
    \centering
    \mbox{\subfigure[]{\includegraphics[width=0.4\textwidth]{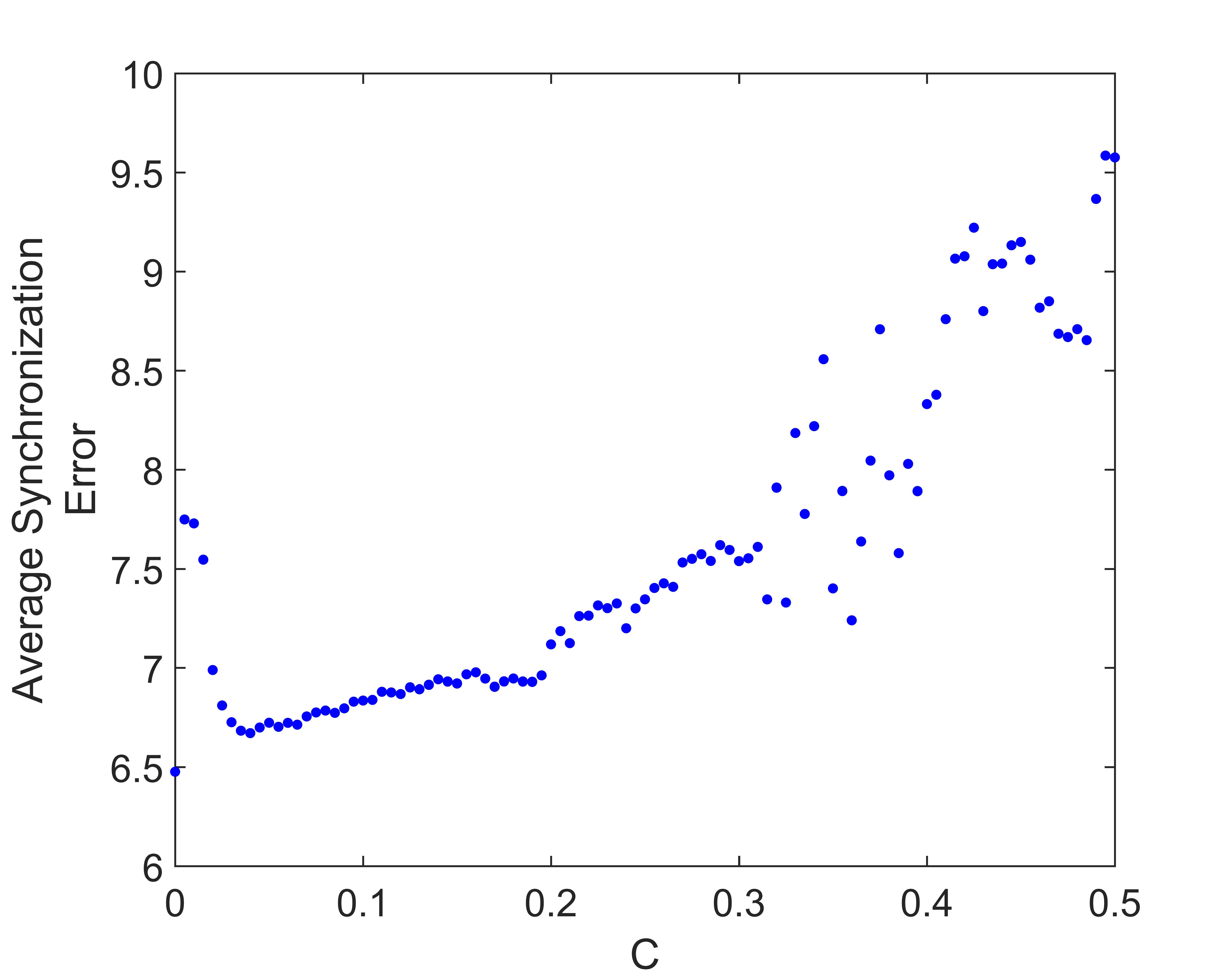}}
    \subfigure[]{\includegraphics[width=0.4\textwidth]{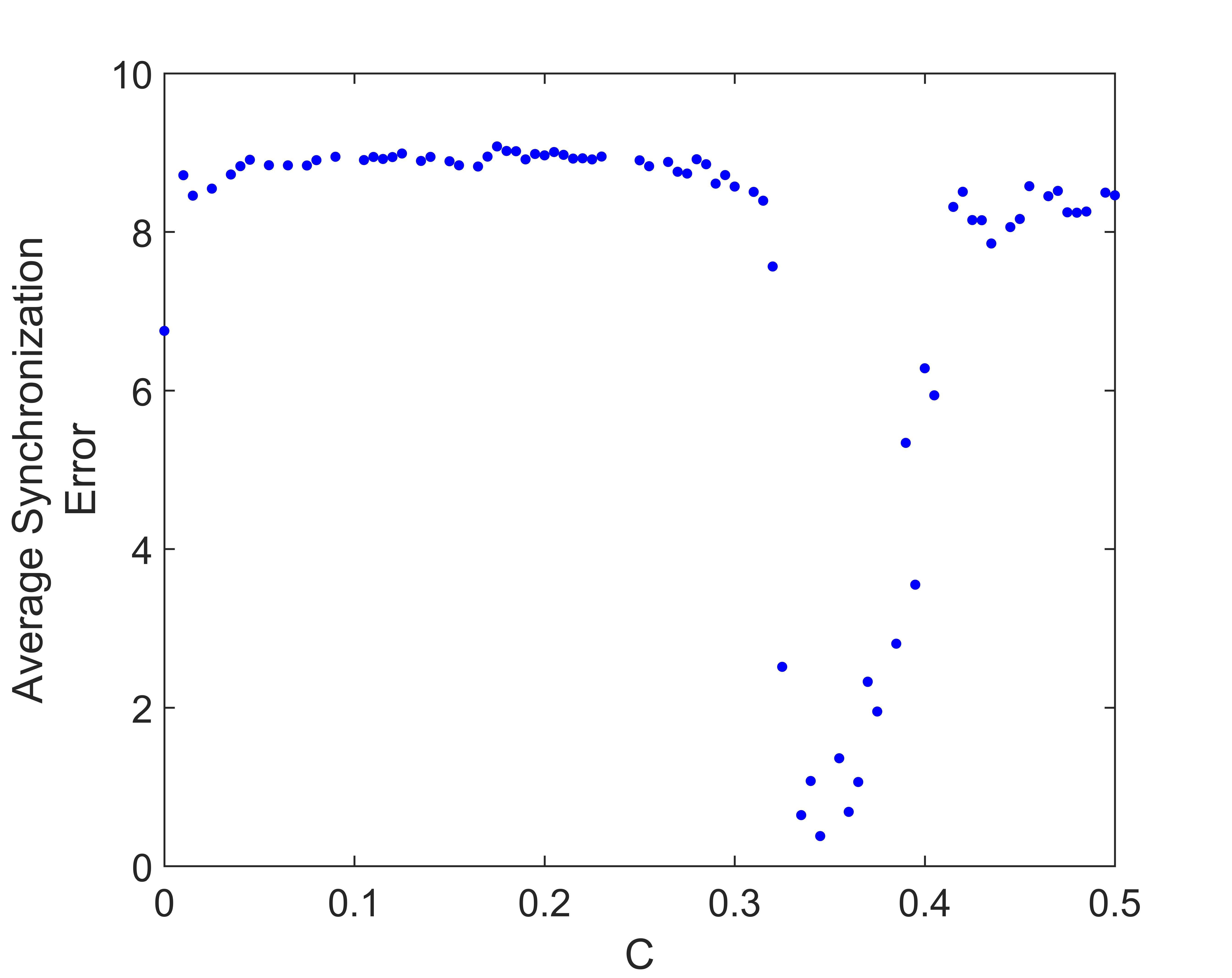}}}
    \caption{Average synchronization error of the coupled patches, with respect to the coupling strength $C$ for Allee parameter (a) $\theta\,=\,0$, (b) $\theta\,=\,0.003$.}
    \label{sync}
\end{figure}

\section*{Extreme Events in the evolution of the population densities}

\begin{figure}[!ht]
    \centering
    \mbox{\subfigure[]{\includegraphics[width=0.33\textwidth]{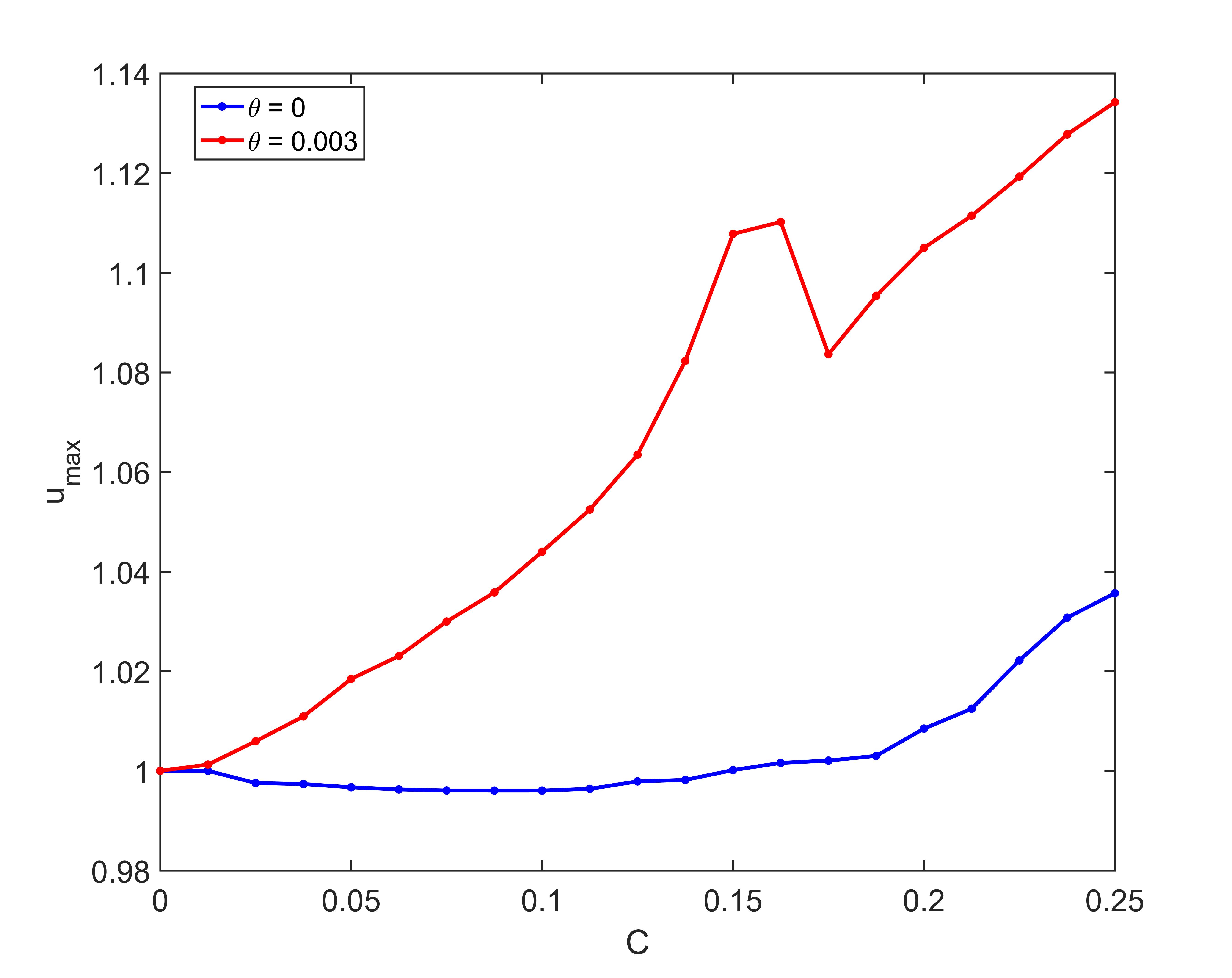}}
    \subfigure[]{\includegraphics[width=0.33\textwidth]{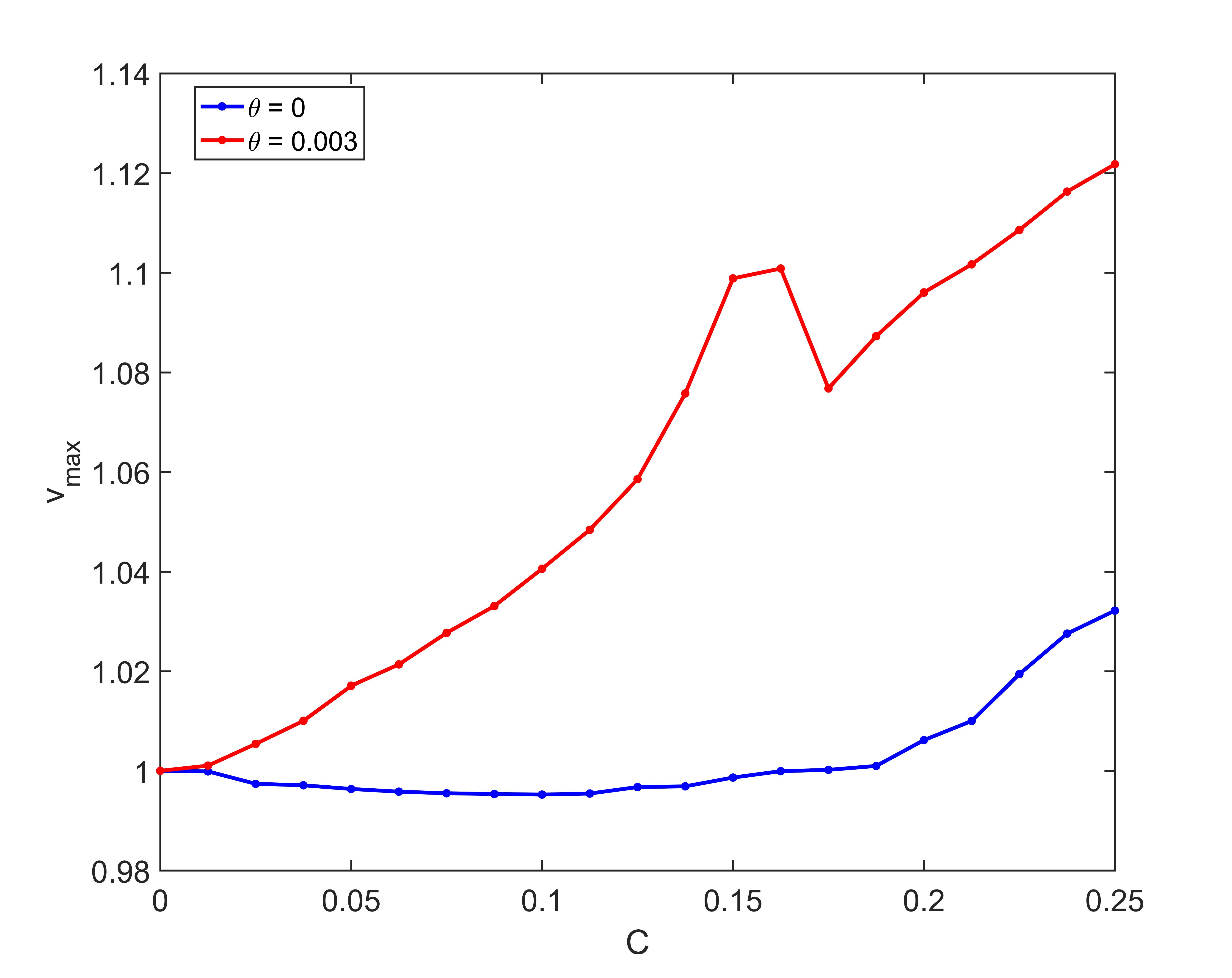}}
    \subfigure[]{\includegraphics[width=0.33\textwidth]{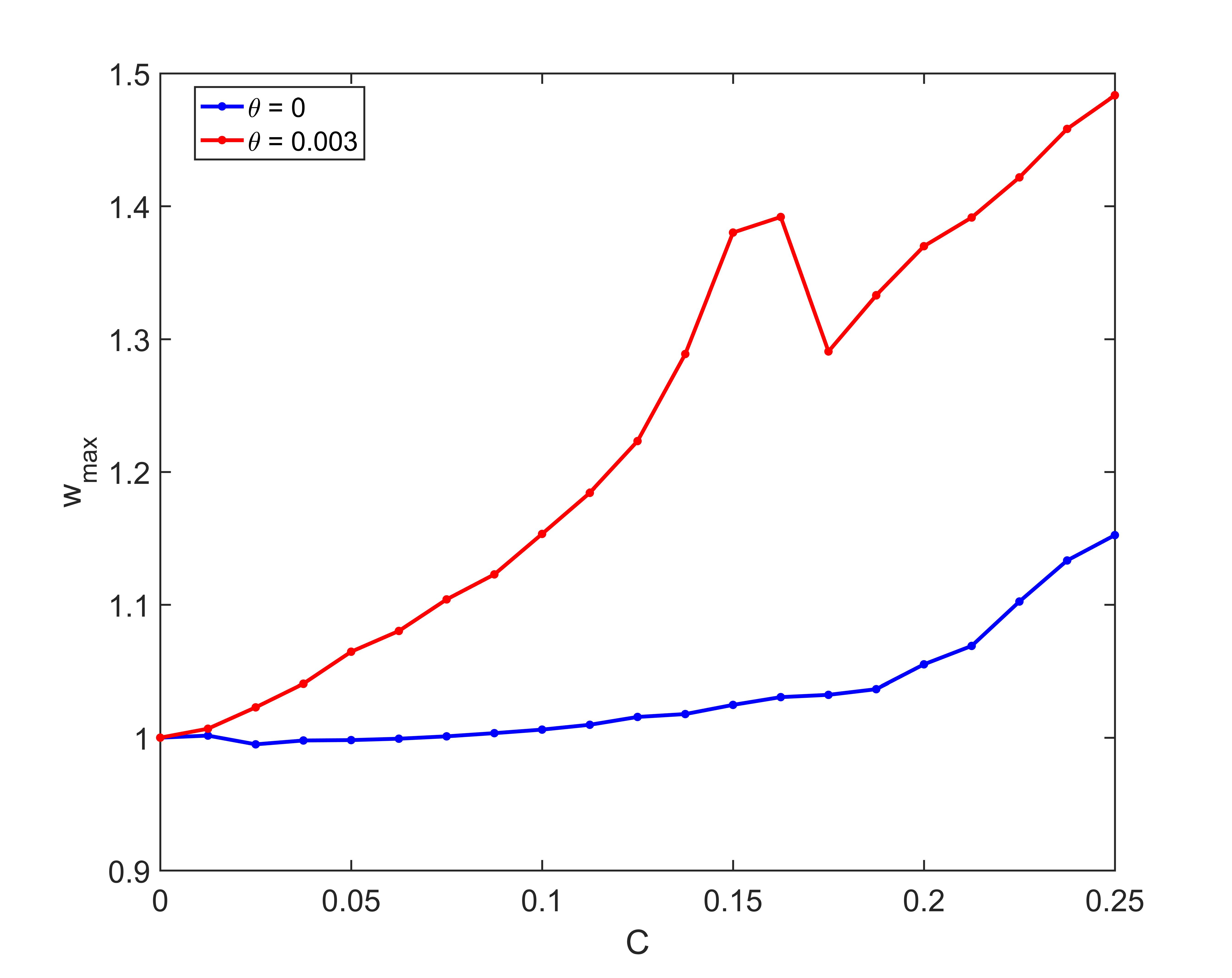}}}
    \caption{Global maximum of (a) vegetation ($u_{max}$), (b) prey population ($v_{max}$) and (c) predator population $(w_{max})$ with respect to the coupling strength $C$, scaled by their values obtained for $C\,=\,0$, for Allee parameter $\theta = 0$ (blue) and $\theta=0.003$ (red). When the value of this scaled quantity is larger than $1$, the global maximum is larger than that for the case of uncoupled patches.}
%    . It is estimated for two different values of $\theta$ with increasing magnitude from a large sample of random initial conditions.}
    \label{fig:globalMax_wrt_coupling}
\end{figure}

We first probe the global maximum of vegetation ($u_{max}$), prey ($v_{max}$) and predator ($w_{max}$) populations under increasing coupling strengths for different Allee parameter $\theta$. We have estimated $u_{max}$, $v_{max}$ and $w_{max}$ by recording the maximum of the population of the patches from a large sample of random initial conditions. In Fig.~\ref{fig:globalMax_wrt_coupling} we plot $u_{max}$,
$v_{max}$ and $w_{max}$ with respect to the coupling parameter $C$ for Allee parameter $\theta= 0$ and $\theta=0.003$. We observe that estimated values of $u_{max}$, $v_{max}$ and $w_{max}$ typically increase with rising coupling strength $C$, and this increase is more pronounced in the presence of Allee effect. This signifies that the Allee effect, as well as coupling strength, enhance the global maximum of the populations of all three species. This observation is consistent with the fact that increasing Allee parameter and coupling strengths typically yield an increase in the size of the attractor in phase space, as evident from the bifurcation diagrams.

%{fig:timeseries&phasep_with_coupling_theta0}e,\,f. 
%Thus by increasing the coupling strength the system moves from a periodic state to chaotic state as for the decoupled system \eqref{eq:finalmodel} when Allee strength being increased.   }

\begin{figure}
    \centering
    \mbox{\subfigure[]{\includegraphics[width=0.4\textwidth]{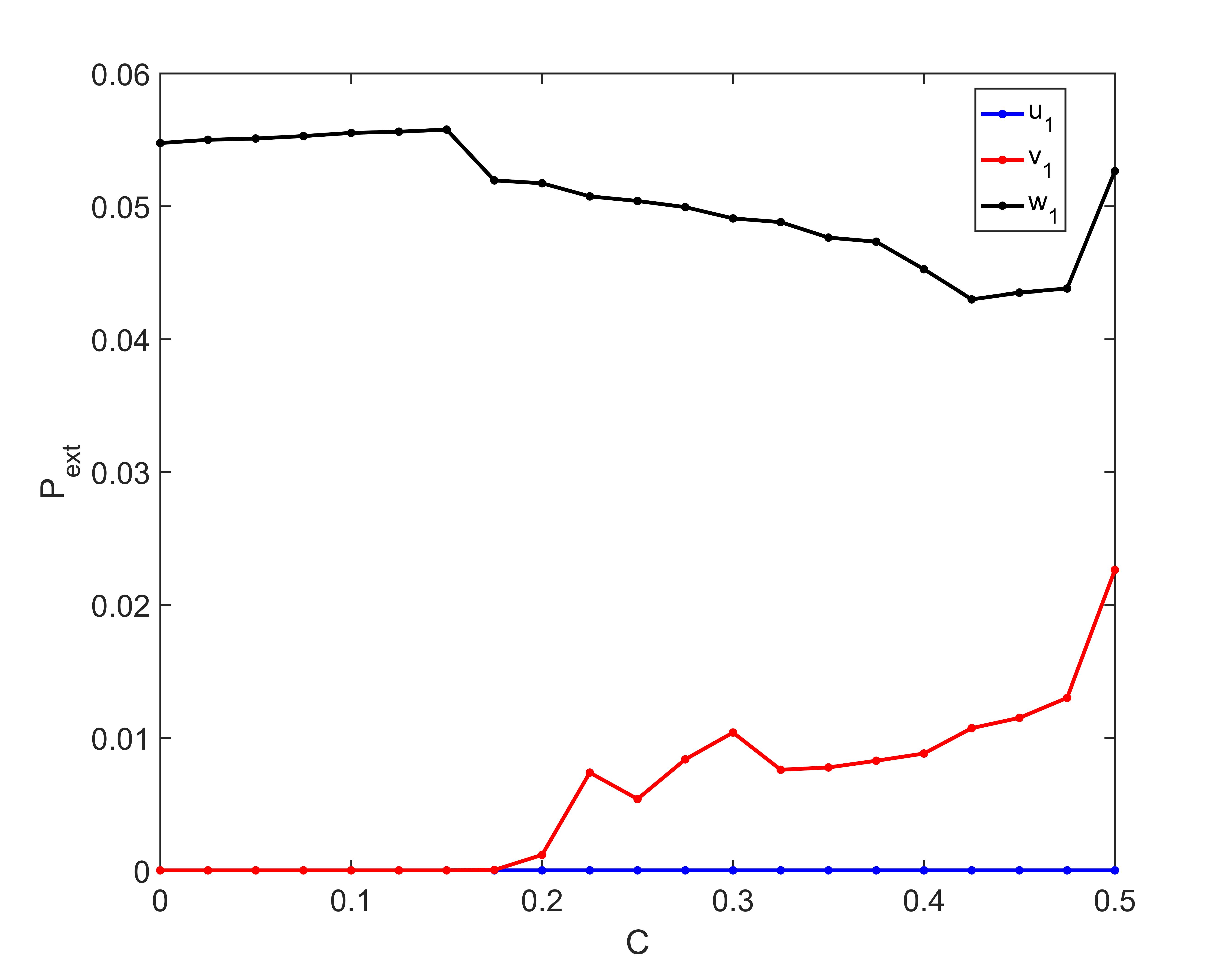}}
    \quad
    \subfigure[]{\includegraphics[width=0.4\textwidth]{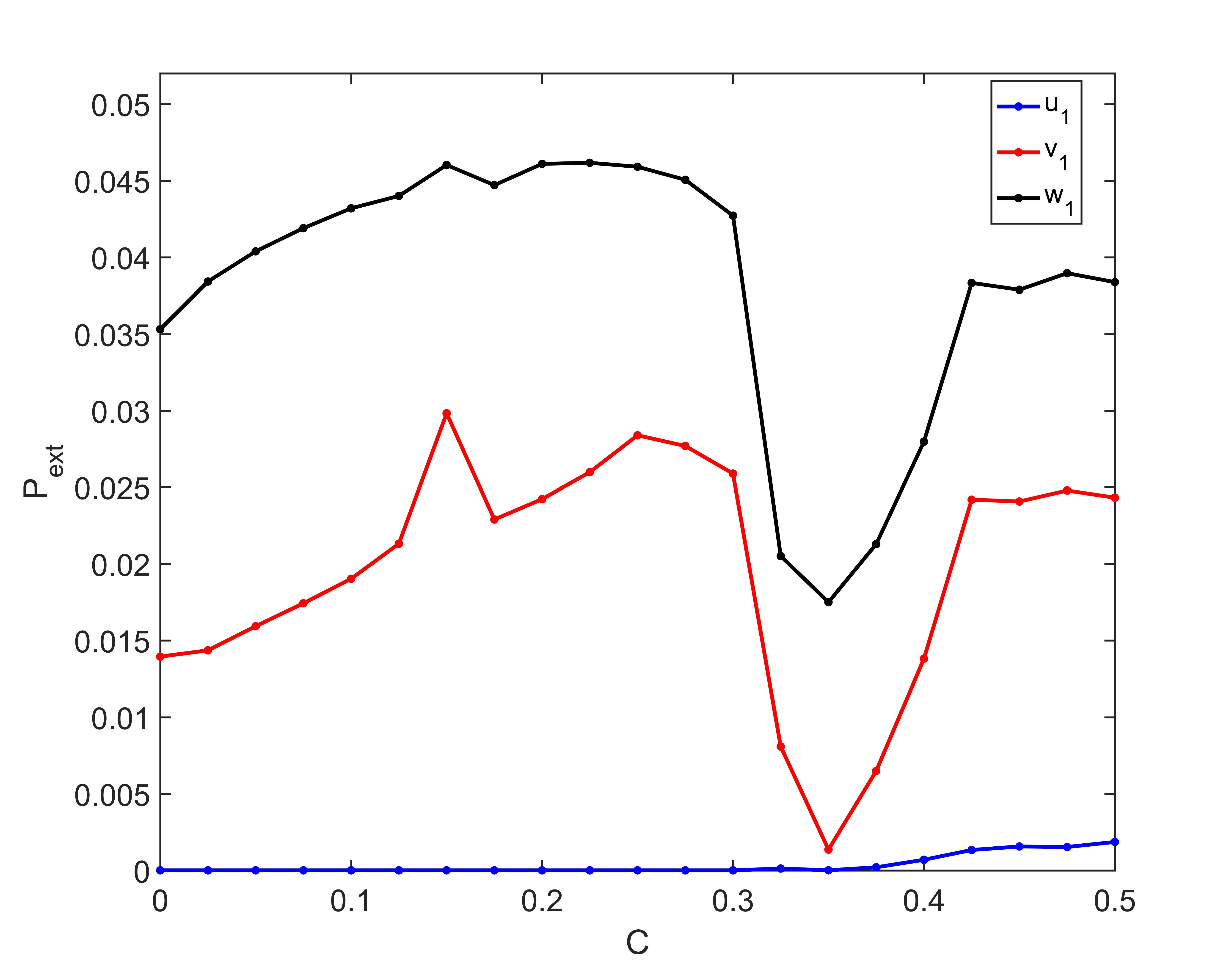}}}
    \caption{Probability of obtaining extreme events in unit time in a patch ($P_{ext}$), with respect to coupling strength $C$, where $P_{ext}$ is estimated by sampling a time interval of length $T\,=\,4000$ and averaging 1000 random initial conditions. We present this for two different values of the Allee parameter: (a) $\theta\,=\,0$ and (b) $\theta\,=\,0.003$. Here we consider that an extreme event occurs when a  population exceeds the threshold value of $\mu + 5\sigma$. The case of vegetation is shown in blue, prey in red and predator in black.}
    \label{fig:no_exteme_wrt_coupling}
\end{figure}

 Next we focus on the probability of obtaining extreme events for vegetation, prey and predator populations in the coupled patches in order to gauge the influence of the Allee effect and the coupling strength on the advent of large deviations. With no loss of generality we consider the threshold of an extreme event to be $5\sigma$. In order to estimate the probability of obtaining extreme events, denoted as $P_{ext}$, we evolve the system from a large sample of random initial conditions over a prolonged time period
 %so that this temporal span is several orders of magnitude longer than the usual oscillation period 
 and note the occurrences when a population exceeds the $\mu + 5\sigma$ threshold. 
 %One noticeable fact is that for example if we look into the time series evolution of the coupled system for $\theta\,=\,0$, $C\,=\,0.5$ (in Fig.~\ref{fig:timeseries&phasep_with_coupling}), prey and predator populations deviate significantly from their mean and cross the prescribed threshold. But this deviation is completely correlated in time and occurs periodically, as the coupled system is confined to a periodic orbit for this choice of parameters. So this cannot be considered as extreme events. 
 %We denote this quantity by $P_{ext}$ and 
 Fig.~\ref{fig:no_exteme_wrt_coupling} displays $P_{ext}$ for all three population densities with respect to the coupling strength for different $\theta$. In general, {\em coupling enhances the occurrences of large deviations.} Note however that when these deviations occur in periodic windows, such near $C\sim 0.5$ for $\theta=0$ and around $C \sim 0.3-0.4$ for $\theta = 0.003$, they are not true extreme events as they are entirely correlated in time and recur periodically.
 
 When Allee effect is absent (Fig.~\ref{fig:no_exteme_wrt_coupling} left panel), vegetation population is always confined to low values and does not show any extreme events. The prey population is likewise limited to small values for lower coupling strengths. However, it deviates considerably from its mean beyond a critical threshold of coupling strength, resulting in extreme events. The predator populations have the highest propensity of extreme events, with extreme events arising even in uncoupled patches. 
 
 In the presence of Allee effect (Fig.~\ref{fig:no_exteme_wrt_coupling} right panel) one finds that both the predator and prey populations display large deviations from the mean, over the full coupling range, including the uncoupled case of $C=0$. The only exception to this trend is the
 %Further, for periodic attractors that are smaller in size, there is corresponding
 perceptible dip in extreme events around $C \sim 0.3-0.4$ for $\theta=0.003$ corresponding to a periodic window supporting a small period-$2$ orbit (cf. Fig.~\ref{fig:Bfdiag_wrt_coupling} lower panels). The most significant result in the coupled patches in the presence of Allee effect is the emergence of a small finite probability of extreme events in the vegetation population for sufficiently strong coupling.
 
\begin{figure}
    \centering
    \includegraphics[width=0.4\textwidth]{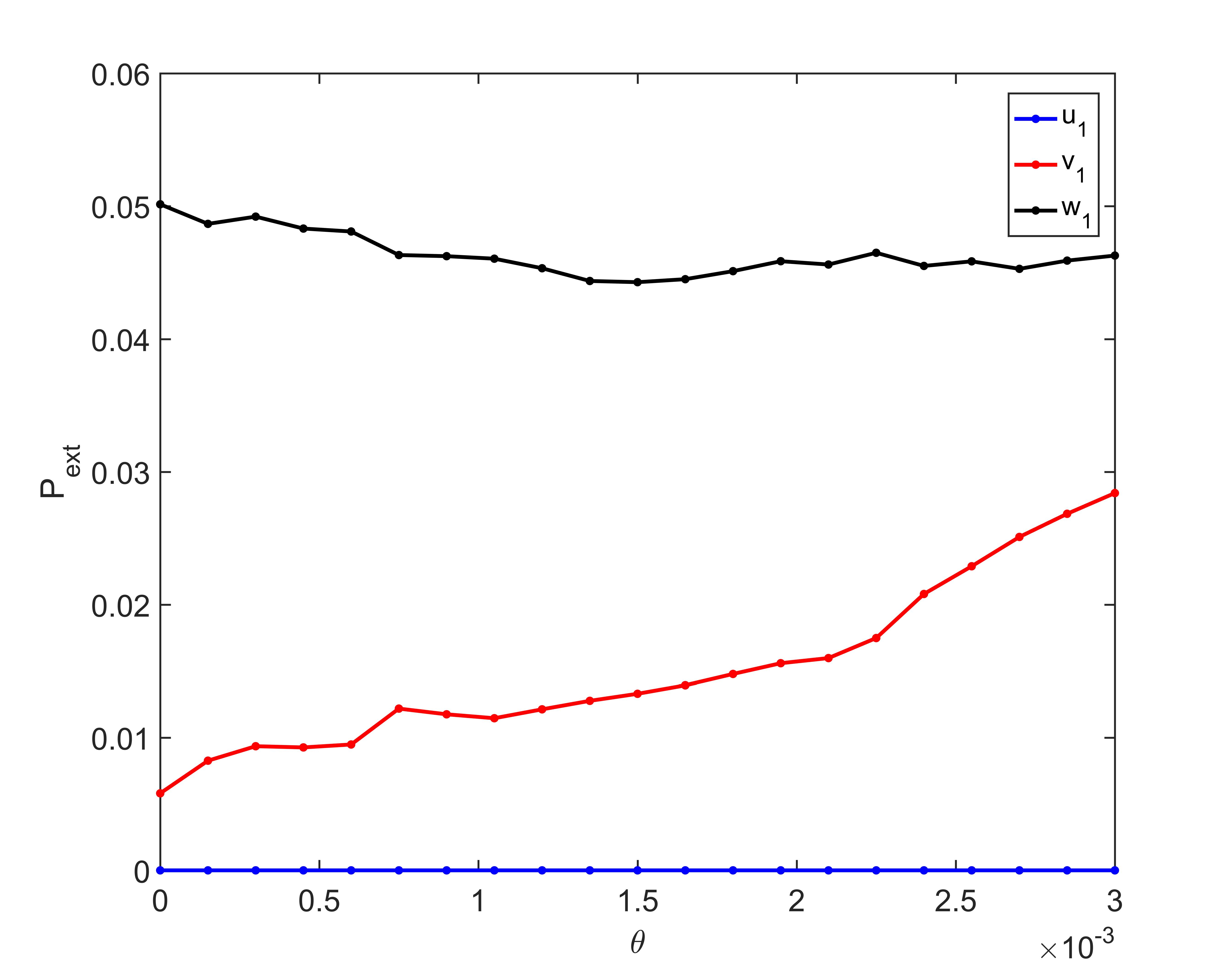}
    \caption{Probability of obtaining extreme events in unit time in a a patch ($P_{ext}$), with respect to Allee parameter $\theta$. Here the coupling strength is fixed at $C\,=\,0.25$, and $P_{ext}$ is estimated by sampling a time interval of length $T\,=\,4000$ and averaging over 1000 random initial conditions. The case of vegetation is shown in blue, prey in red and predator in black.}
    \label{fig:no_extreme_wrt_theta_coupling0dot5}
\end{figure}

 These trends are further bourne out by Fig.~\ref{fig:no_extreme_wrt_theta_coupling0dot5}, which displays $P_{ext}$ over a range of Allee parameters for a fixed value of coupling strength. It is again apparent that predator populations exhibit the highest propensity for large deviations from the mean, and this is not affected much by the magnitude of the Allee Effect. The prey population on the other hand shows steady increase in extreme events with increasing Allee parameter. However, $P_{ext}$ for the prey is always lower than that for predators. The vegetation shows no extreme events at this value of coupling, even when the Allee effect is present.

\section*{Effect of Noise on Extreme Events}

Noise is prevalent in natural ecosystems as a result of external factors such as inherent diversity, fluctuations in migration, and environmental changes. Here we examine how stochasticity influences the three-species coupled system \eqref{eq:coupled_system}. Specifically  we explore the dynamics of the system under additive random noise $\xi(t)$, given by the following dynamical equations:

\begin{equation}
\label{eq:model_with_additive_noise}
    \begin{split}
        \dot{u}_{1} &= f(u_{1},v_{1},w_{1}) \ + \ \xi_{1}(t),\\
        \dot{v}_{1} &= g(u_{1},v_{1},w_{1}) \ - \ C \ v_{1}w_{2} \ + \ \xi_{2}(t),\\
        \dot{w}_{1} &= h(u_{1},v_{1},w_{1}) \ + \ C \ v_{2}w_{1} \ + \ \xi_{3}(t),\\
         \ & \ \\
        \dot{u}_{2} &= f(u_{2},v_{2},w_{2}) \ + \ \xi_{4}(t),\\
        \dot{v}_{2} &= g(u_{2},v_{2},w_{2}) \ - \ C \ v_{2}w_{1} \ + \ \xi_{5}(t),\\
        \dot{w}_{2} &= h(u_{2},v_{2},w_{2}) \ + \ C \ v_{1}w_{2} \ + \ \xi_{6}(t).
    \end{split}
\end{equation}

%\begin{figure}
%    \centering
%    \subfigure[]{\includegraphics[width=0.49\textwidth]{PhaseP_patch1_theta0dot002_C0dot5.png}}}
%    \mbox{\subfigure[]{\includegraphics[width=0.49\textwidth]{TimeSeries_withNS_patch1_theta0dot002_C0dot5_sig0dot0001.png}}
%    \subfigure[]{\includegraphics[width=0.49\textwidth]{PhaseP_withNS_patch1_theta0dot002_C0dot5_sig0dot0001.png}}}
%    \mbox{\subfigure[]{\includegraphics[width=0.49\textwidth]{TimeSeries_withNS_patch1_theta0dot002_C0dot5_sig0dot001.png}}
%    \subfigure[]{\includegraphics[width=0.49\textwidth]{PhaseP_withNS_patch1_theta0dot002_C0dot5_sig0dot001.png}}}
%    \mbox{\subfigure[]{\includegraphics[width=0.49\textwidth]{TimeSeries_withNS_patch1_theta0dot002_C0dot5_sig0dot01.png}}
%    \subfigure[]{\includegraphics[width=0.49\textwidth]{PhaseP_withNS_patch1_theta0dot002_C0dot5_sig0dot01.png}}}
%    \caption{Time series for the vegetation, prey and predator populations and corresponding phase attractor of the system, for coupling strength $C\,=\,0.5$, Allee parameter $\theta\,=\,0.002$ and noise strength $\sigma\,=\,0$ (a-b), $\sigma\,=\,10^{-4}$ (c-d), $\sigma\,=\,10^{-3}$ (e-f) and $\sigma\,=\,10^{-2}$ (g-h).}
%    \label{fig:timeseries&phasep_coupling_withaddNS}
%\end{figure}

\begin{figure}
    \centering
    \mbox{\subfigure[]{\includegraphics[width=0.42\textwidth]{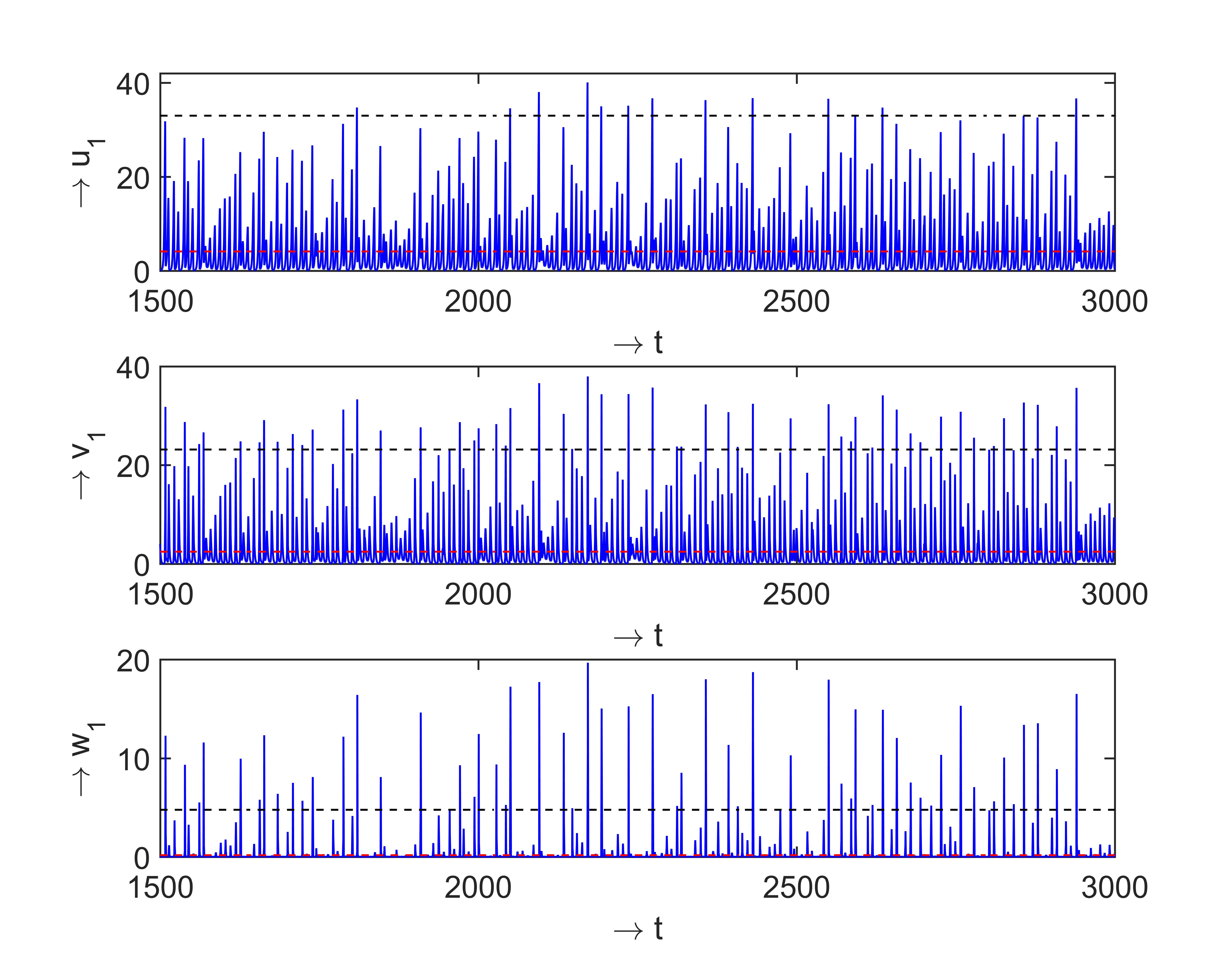}}
    \subfigure[]{\includegraphics[width=0.42\textwidth]{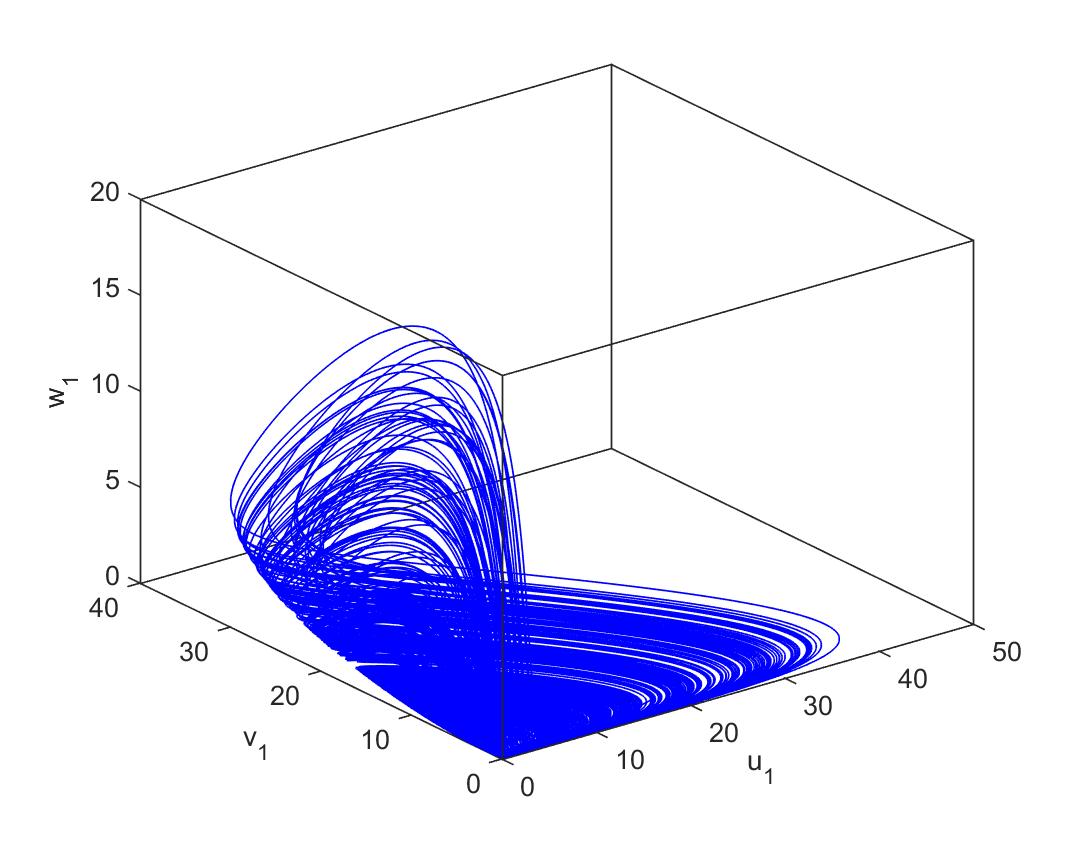}}}
    \mbox{\subfigure[]{\includegraphics[width=0.42\textwidth]{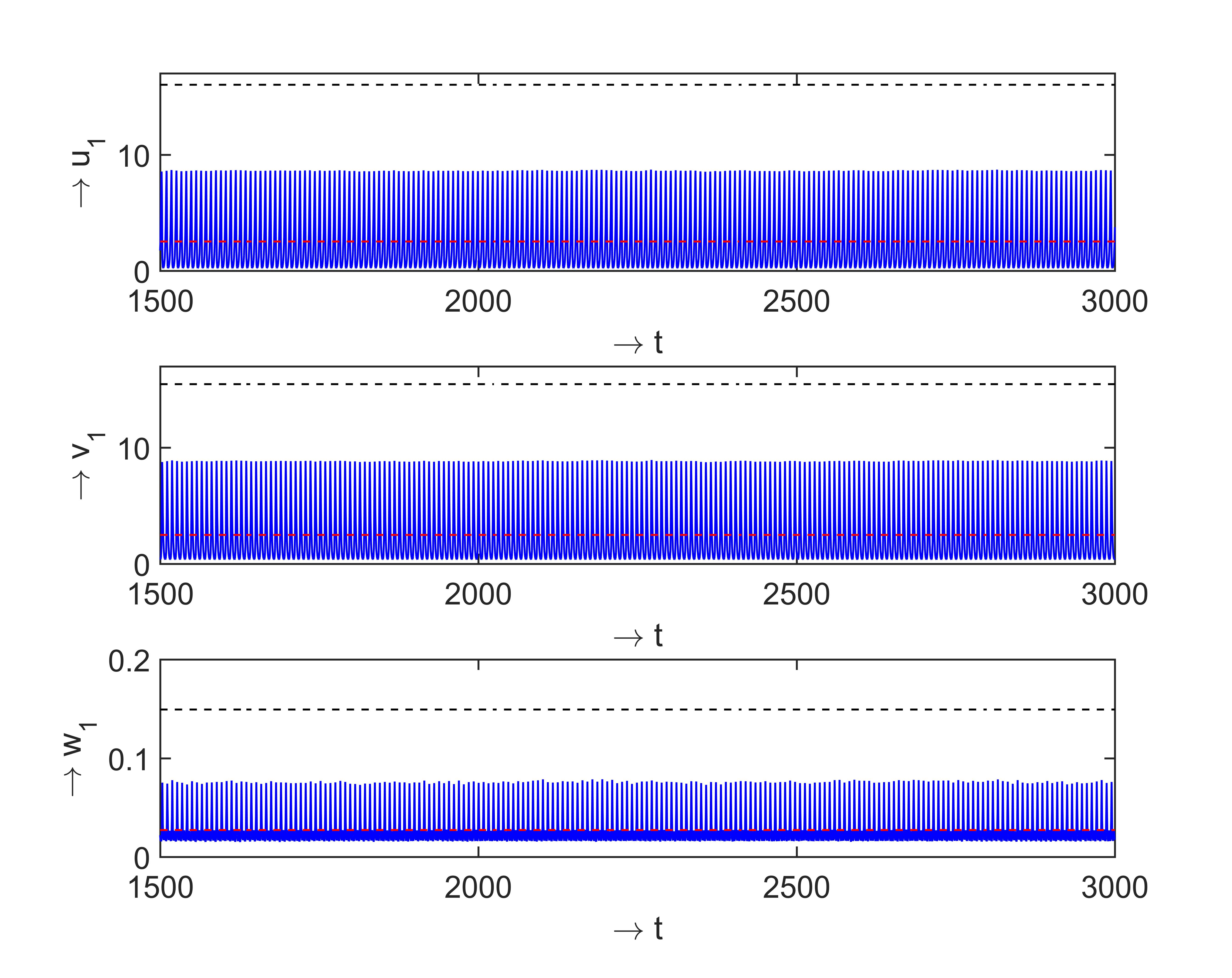}}
    \subfigure[]{\includegraphics[width=0.42\textwidth]{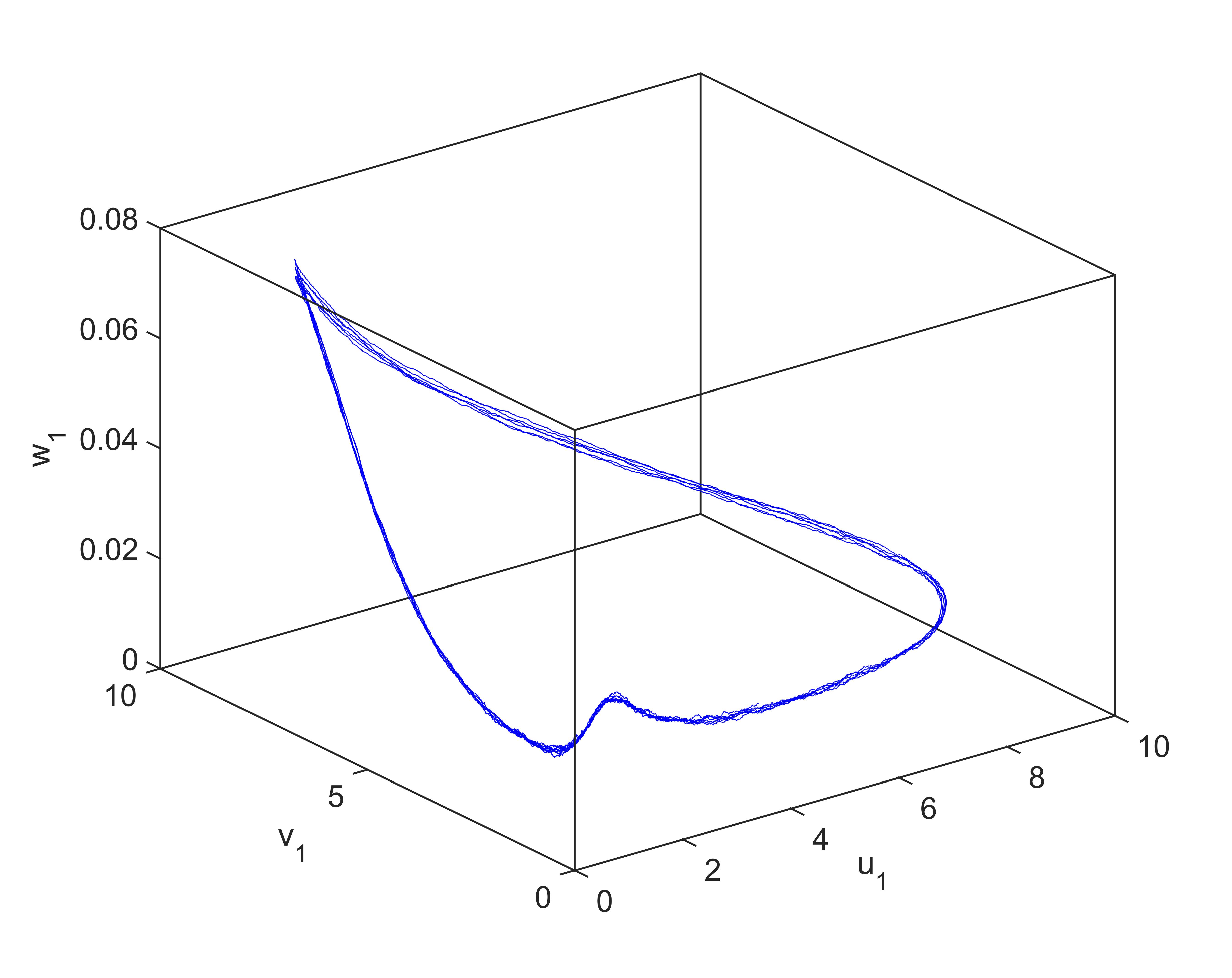}}}
    \mbox{\subfigure[]{\includegraphics[width=0.42\textwidth]{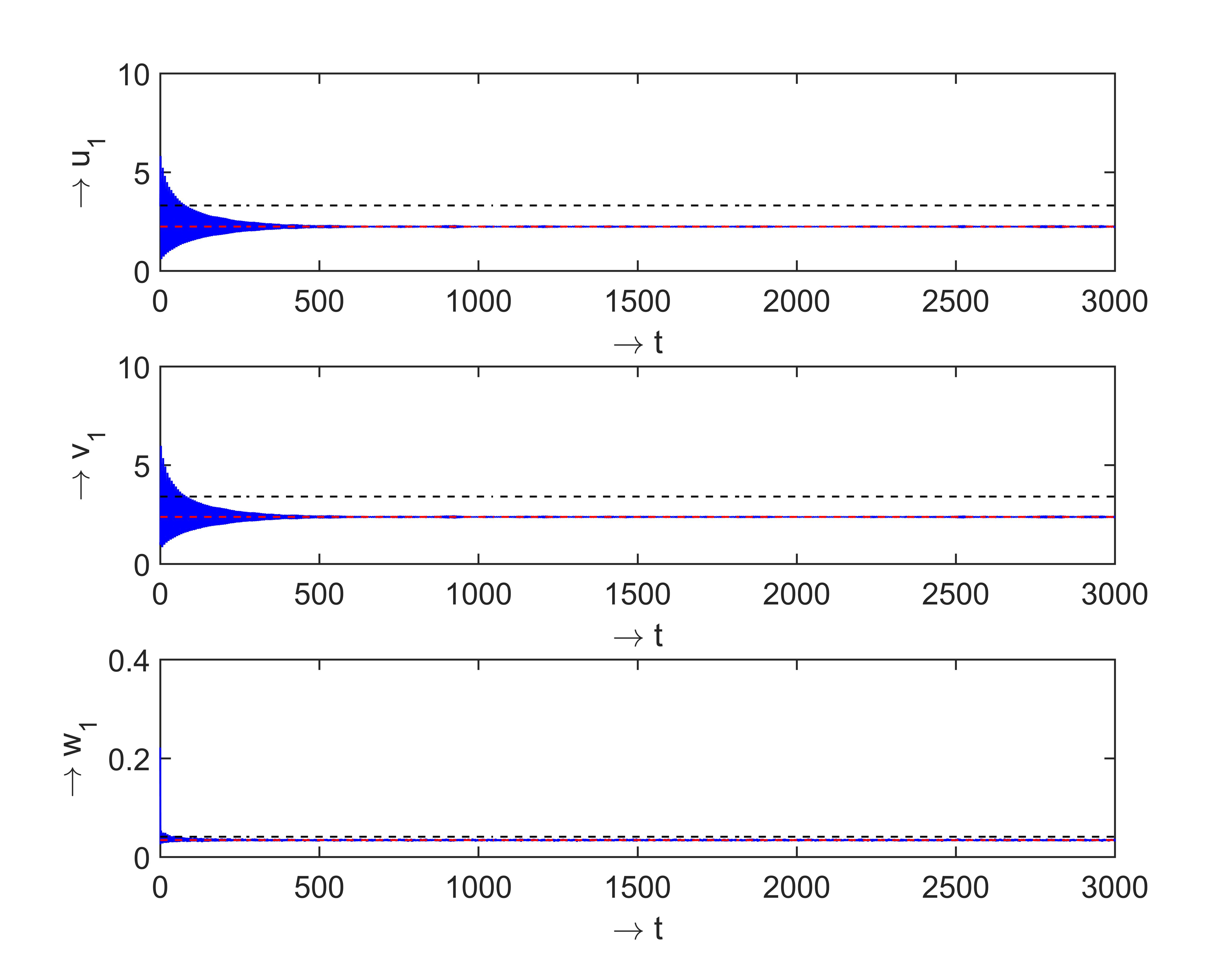}}
    \subfigure[]{\includegraphics[width=0.42\textwidth]{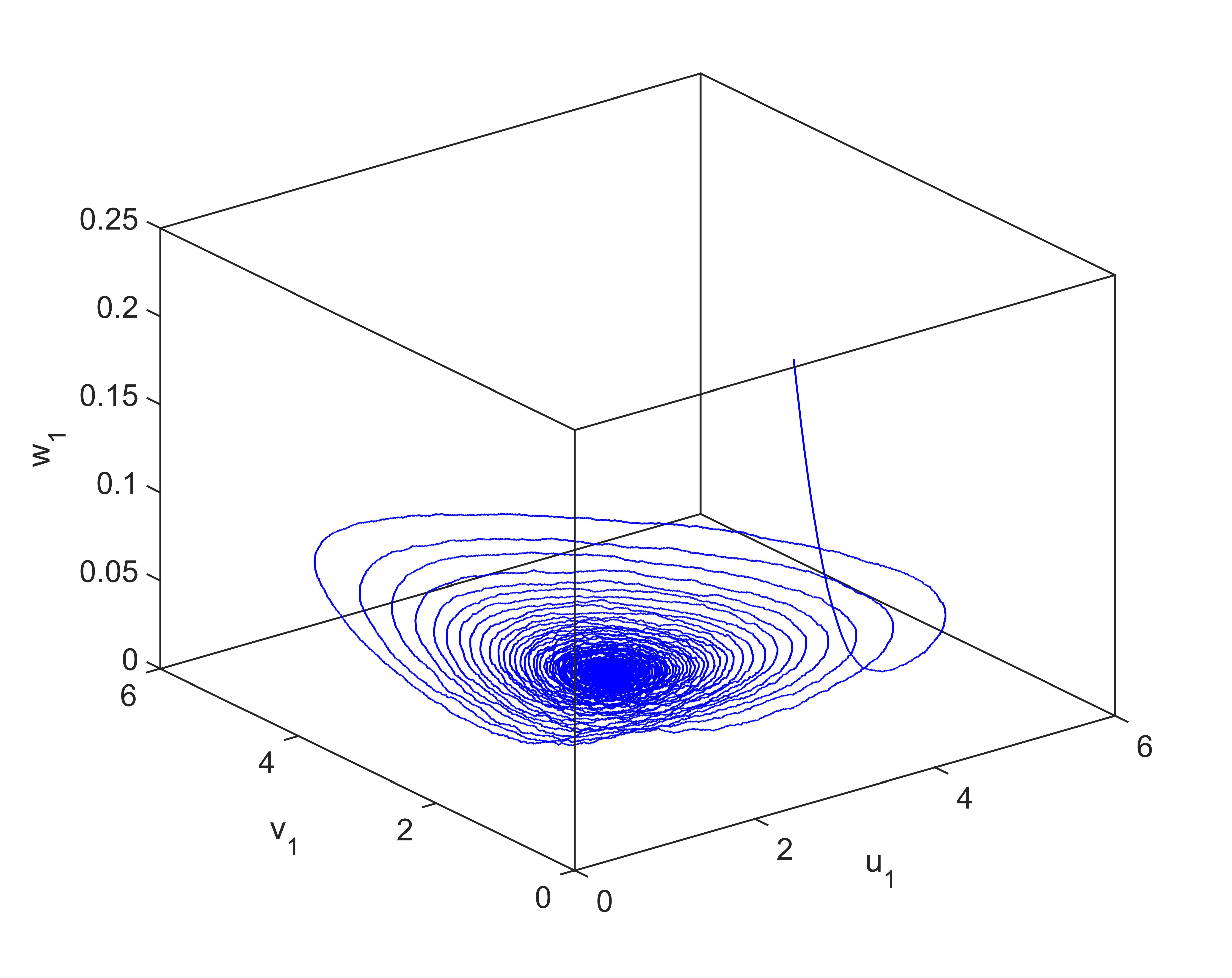}}}
    \caption{Time series for the vegetation, prey and predator populations (left panel) and corresponding phase attractor of the system (right panel), for coupling strength $C\,=\,0.5$, Allee parameter $\theta\,=\,0.05$ and noise strength (a-b), $\sigma\,=\,0.001$ (c-d), $\sigma\,=\,0.005$ (e-f) $\sigma\,=\,0.01$. Note that for this value of $C$ and $\theta$, the system has unbounded vegetation growth in the absence of noise. So noise has suppressed the explosive growth in the coupled system.}
    \label{fig:timeseries&phasep_coupling_withaddNS_theta0dot05}
\end{figure}

\medskip
\noindent The functional forms $f (u,v,w)$, $g (u,,w)$ and $h(u,v,w)$ are given as before in Eq.~\ref{eq:finalmodel}, and $\xi_{i}(t),\,i\,=\,1,2,\cdots,6$ represent zero mean delta-correlated Gaussian white noises satisfying  $<\xi_{i}(t),\xi_{j}(t^{'})>\,=\,\sigma \delta(t-t^{'})\delta_{ij}$ for $i,j\,=\,1,2,\cdots,6$, where $\sigma$ is the noise strength.
We investigate the dynamics of the stochastic differential system \eqref{eq:model_with_additive_noise} numerically by using explicit Euler-Maruyama scheme. We have also checked the convergence and stability of the numerical solutions with decreasing step size. 

We investigate how the dynamics of the coupled system is affected by the presence of additive noise. In particular, we explore whether the system becomes more irregular, or if chaos is suppressed to noisy cycles or noisy fixed points, in the presence of stochasticity. In Fig.~\ref{fig:timeseries&phasep_coupling_withaddNS_theta0dot05}
%{fig:timeseries&phasep_coupling_withaddNS} 
we present the time series and phase portrait of the system \eqref{eq:model_with_additive_noise} for different noise strengths. It is observed that when noise strength is either zero or of very low magnitude ($\sigma\,=\,10^{-3}$), all three populations oscillate aperiodically and system evolves to chaotic attractors (Fig.~\ref{fig:timeseries&phasep_coupling_withaddNS_theta0dot05}
%{fig:timeseries&phasep_coupling_withaddNS}
(a-b)
%,(c-d)
). Also note that populations of all species exhibit extreme events for low magnitude of the noise strength as their states occasionally cross the $\mu + 5\sigma$ threshold. On increasing noise strengths ($\sigma \sim 5 \times 10^{-3}$) , these extreme events completely disappear from the all populations, and the populations of all three species  settle into a noisy periodic orbit (Fig.~\ref{fig:timeseries&phasep_coupling_withaddNS_theta0dot05}
%{fig:timeseries&phasep_coupling_withaddNS}
%(e-f)
(c-d)). On further increasing the noise strength ($\sigma \sim 10^{-2}$), the populations of all three species settle down to a noisy quasi-steady state (Fig.~\ref{fig:timeseries&phasep_coupling_withaddNS_theta0dot05}
%{fig:timeseries&phasep_coupling_withaddNS}(g-h)
(e-f)), and the system continues to exhibit no extreme deviations from the mean. So then, very interestingly, the extreme events in the populations of this coupled three-species system are suppressed under sufficiently strong noise. Therefore we arrive at the following significant conclusion: additive noise not only suppresses the extreme events from the vegetation, prey and predator populations, but it also transforms the dynamics of the system from chaos to a noisy quasi-steady state. Additionally the additive noise also tames the explosive growth in the vegetation population that we observed earlier.

\section{Discussion}

In summary, we explored the dynamics of two coupled patches of a three-species trophic system incorporating the Allee Effect in the prey population. Our focus is on the emergence of extreme events in the system. In particular we address the significant question of whether or not Allee effect and coupling suppress or enhance extreme events. 
%Further we consider the influence of stochasticity on extreme events. 
Our first key observation is as follows:
%: First, under Allee effect the regular periodic dynamics changes to chaotic, as evident from the emergence of chaotic attractors for increasing Allee parameter $\theta$. Further, 
we find that the system experiences a explosive blow-up in the vegetation population after a critical value of coupling strength in presence of the Allee effect. 
%The most significant result is the observation of a critical Allee parameter beyond which the probability of obtaining extreme events becomes non-zero for all three population densities. Though the emergence of extreme events in the predator population is not affected much by the Allee effect, the prey population shows a sharp increase in the probability of obtaining extreme events after a threshold value of the Allee parameter $\theta$, and the vegetation population also yields extreme events for sufficiently strong Allee effect. An interesting open problem in this context would be to check the observation that the extreme events in the predator population are more pronounced than in prey and vegetation across other models, in order to establish the generality of this important trend in a larger class of models. 
Further the interplay of the Allee effect and coupling has a pronounced influence on the nature of the dynamics. In order to explore this aspect in detail, we looked into the bifurcation scenarios of the system with respect to the Allee parameter $\theta$ and coupling strength $C$. We find that the populations of all three species of the coupled system \eqref{eq:coupled_system} oscillate in a regular manner and settle into a four periodic orbit when the Allee parameter $\theta$ is low, whereas, the populations fluctuate aperiodically and chaotic attractors emerge in the coupled system with increasing the Allee parameter $\theta$.
%through a period-doubling cascade. 
In addition the size of the attractor gradually increases with the increasing Allee parameter $\theta$ which is compatible with trends from a single patch. %We also demonstrate bifurcation diagrams of all populations with respect to the coupling strength in Fig.~\ref{fig:Bfdiag_wrt_coupling} for two different values of $\theta$. 
It was also clearly evident that coupling induces chaotic behavior in the system.
%, even when Allee effect is absent. Further, one also observes periodic windows arising as a result of interior crisis. Similar trends of chaotic regimes, 
Further, interspersed in the chaotic regimes one finds periodic windows that arise from interior crisis in certain ranges of coupling strengths.
%, can be seen in the bifurcation diagrams when the Allee parameter is non-zero. 
The notable difference stemming from the Allee effect is that there is chaos for low coupling strengths, including the case of $C=0$ (i.e. the uncoupled case) for finite $\theta$, while weakly coupled patches with no Allee effect exhibit regular dynamics. 

Most importantly, we observe that for large enough coupling strengths and Allee parameters all population densities exhibit non-zero  probability of yielding extreme events. In general, the predators have the largest propensity for extreme events in coupled patches, and the vegetation population densities exhibit the least number of extreme occurrences. 
%for all values of the Allee parameter $\theta$ (inclusive of $\theta = 0$).
Further, the emergence of extreme events in the predator population is not affected much by either the coupling strength or the Allee effect. For prey populations, in the absence of the Allee effect there are no extreme events for low coupling strengths, but there is a sharp increase in extreme events after a critical value of coupling strength. For the vegetation population a small finite probability of extreme events emerges for strong enough coupling, only in the presence of Allee effect. 

Lastly we consider the influence of additive noise on extreme events. First, we find that noise tames the unbounded vegetation growth induced by the coupling and Allee effect. More interestingly, we demonstrate that stochasticity drastically diminishes the probability of extreme events in all three populations in the coupled atches. In fact for sufficiently high noise, we do not observe any more extreme events in the system. This indicates that noise can mitigate extreme events, and has potentially important impact on the observability of extreme events in naturally occurring systems.\\
 
%1) Due to the Allee effect the dynamics of this three-species model changes from regular to chaotic, as evident from the emergence of chaotic attractors for increasing $\theta$.

%2) For single patches, it is evident that there is a critical $\theta$ after which the probability of obtaining extreme events becomes non-zero for vegetation.

%3) For single patches, prey shows sharp increase in the probability of obtaining extreme events with increasing $\theta$.

%4) The probability of obtaining extreme events in predator populations is not as significantly affected by the Allee effect. There is a decrease in extreme events for increasing $\theta$, but this is relatively small.

%Further even for high coupling strengths ($C\sim 1$) we do not obtain synchronization of the populations in the patches.

%In coupled population patches the probability of obtaining extreme events increases with coupling strength, in presence of Allee effect.

\section*{Acknowledgement}

This paper is in honor of Somdatta Sinha, who continues to inspire us to use mathematical formalisms to understand wide-ranging problems arising in biological contexts.

\bibliographystyle{abbrv}
\bibliography{ref}

%\printbibliography
\end{document}